\documentclass{article}
\usepackage{fullpage}
\usepackage{natbib}

\usepackage{amssymb,amsmath,graphicx,bm}
\def\e{\mathrm{e}}

\def\d{\mathrm{d}}

\def\eps{\epsilon}

\def\E{{\mathbb{E}}}

\def\bx{\bm{x}}
\def\by{\bm{y}}

\def\bu{\bm{u}}
\def\bv{\bm{v}}

\def\boeta{\bm{\eta}}

\def\bX{\bm{X}}

\def\ip{\lrcorner}

\def\bsigma{\bm{\sigma}}

\newcommand{\eqn}[1]{(\ref{eq:#1})}
\newcommand{\lab}[1]{\label{eq:#1}}
\newcommand{\inter}[1]{\quad \textrm{#1} \quad}

\def\dcc{d_{\textsc{cc}}}

\def\beq{\begin{equation}}
\def\eeq{\end{equation}}
\newcommand\ch[1]{#1}

\title{{\Large \textbf{Ocean neutral transport: sub-Riemannian geometry and hypoelliptic diffusion}}}

\author{Matthieu Chatelain$^\dag$, Isambard Goodbody$^\ddag$, Nived Rajeev Saritha$^\ddag$ and Jacques Vanneste$^{\ddag}$\thanks{Corresponding author. Email: j.vanneste@ed.ac.uk}}

\date{
$\dag$Laboratoire de Physique (UMR CNRS 5672), ENS de Lyon, % 46 All\'ee d'Italie, 
 F-69364 Lyon cedex 07, France
 \\ 
$\ddag$School  of  Mathematics  and  Maxwell  Institute  for  Mathematical  Sciences,  University  of  Edinburgh, Edinburgh EH9 3FD, UK}
\begin{document}

\maketitle

\begin{abstract}
\noindent
\ch{Transport and mixing of tracers} in the ocean is thought to be preferentially along neutral planes defined by  \ch{the} potential temperature and salinity fields. This gives rise to a conceptual model of ocean transport in which \ch{water} parcel trajectories are everywhere neutral, that is, tangent to the neutral planes. Because the distribution of neutral planes is not integrable, neutral transport, while locally two dimensional, is globally three dimensional. We describe this form of transport, building on its connection with contact and sub-Riemannian geometry. 

We discuss a Lie-bracket interpretation of 
local dianeutral transport, the quantitative meaning of helicity and   the implications of the accessibility theorem. We compute sub-Riemnanian geodesics for climatological neutral planes and put forward the use of the associated Carnot--Carath\'eodory distance as a diagnostic of the strong anisotropy of neutral transport. 

We propose a stochastic toy model of neutral transport  which represents motion along neutral planes by a Brownian motion. The corresponding diffusion process is degenerate and not (strongly) elliptic. The non-integrability of the neutral planes however ensures that the diffusion is hypoelliptic. As a result, trajectories are not confined to surfaces but visit the entire three-dimensional ocean. 
The short-time behaviour is qualitatively different from that obtained with a non-degenerate highly anisotropic diffusion. We examine both short- and long-time behaviours using Monte Carlo simulations. The simulations provide an estimate for the time scale of ocean vertical transport implied by the  constraint of neutrality.

\end{abstract}

\section{Introduction}

The motion of water parcels in the ocean is approximately two-dimensional as a result of  strong density stratification and  weakness of diabatic processes. The simplest model for this assumes that fluid parcels conserve their density. Their motion is then confined to isopycnal surfaces. Matters are more complicated when salinity and the nonlinearity of the equation of state for seawater are taken into account. \citet{mcdo87} argues that the \ch{velocity of water parcels is} everywhere tangent to a family of planes -- the neutral planes -- determined by \ch{the} potential temperature and salinity distribution of the ocean. 

%The direction of the neutral plane at each location $\bx$ encoded in $\boeta$ is determined by a physical argument.
\citet{mcdo87} defines neutral planes by requiring that the density of water parcels that are displaced along them adiabatically, that is, conserving their entropy and salinity, matches the density of the parcels they replace. Such displacements  leave the potential energy unchanged. For an equation of state of the form
\beq
\rho = R(\theta,S,p),
\eeq
where $\rho$ is the density, $\theta$ the potential temperature (\ch{used in place of} entropy), $S$ the salinity and $p$ the pressure, 
a parcel displaced in this way from position $\bx$ to position $\bx + \eps \bu$ for $\eps \ll 1$ has density $R\left(\theta(\bx),S(\bx),p(\bx + \eps \bu)\right)$. Equating this with the local density at $\bx + \eps \bu$ gives the condition
\beq
R\left(\theta(\bx),S(\bx),p(\bx + \eps \bu)\right) = R\left(\theta(\bx + \eps \bu),S(\bx + \eps \bu)),p(\bx + \eps \bu)\right).
\lab{constdens}
\eeq
On Taylor expanding, this reduces to
\beq
\bm{n} \cdot \bu  = 0,
\lab{perp}
\eeq
where
\beq
\bm{n} \propto R_\theta  \, \nabla \theta + R_S \,  \nabla S
\lab{neutralvec}
\eeq
defines the dianeutral vector, perpendicular to the neutral planes.

In \eqn{neutralvec} the subscripts denote partial derivatives and  $\propto$ denotes equality up to multiplication by a scalar field. Such a multiplication is clearly irrelevant for the definition of the neutral planes -- many authors, starting with \citet{mcdo87}, use $\bm{n} = \beta \, \nabla S - \alpha \, \nabla\theta$, where $\beta = R_S/R$ and $\alpha= - R_\theta/R$ are the coefficients of saline contraction and thermal expansion.

\begin{figure}
\begin{center}
\includegraphics[height=4cm]{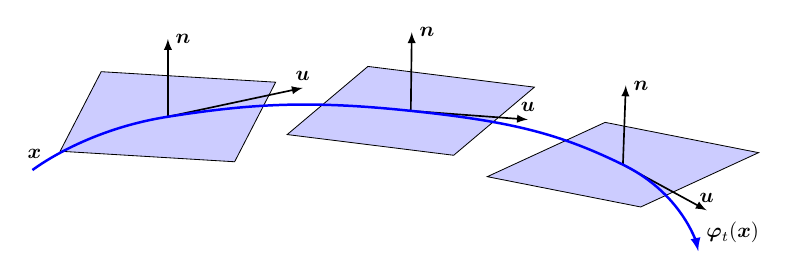}
\caption{Distribution of neutral planes defined by a field of normal $\bm{n}$. A neutral path, with $\bx$  the initial position and $\bm{\varphi}_t(\bx)$ the position at time $t$, is shown as the blue curve. }
\label{fig:neutralplanes}
\end{center}
\end{figure}

Eq.\ \eqn{neutralvec} defines a field of neutral vectors and neutral planes, usually computed from time-independent potential temperature and salinity fields obtained e.g.\ from climatology data.
The path of a water parcel is then said to be neutral if it is everywhere tangent to the neutral planes so that its velocity $\bu$  satisfies \eqn{perp}. See figure \ref{fig:neutralplanes} for an illustration.

Much of the work on neutral motion focuses on neutral surfaces and neutral density coordinates. Neutral surfaces are surfaces that are  everywhere approximately tangent to the neutral planes. As pointed out by \citet{mcdo-jack88}, the tangency can only be approximate because the neutral planes fail to satisfy the integrability condition
$\bm{n} \cdot \nabla \times \bm{n} = 0$ \citep[see also][and \S\ref{sec:noninte} below]{mcdo-jack97}. The failure is suitably small \citep{mcdo-jack07} so that useful surfaces that are approximately neutral can be constructed.
A neutral density is a variable that labels these (approximate) neutral surfaces. It can be used as a vertical coordinate.
Neutral surfaces and neutral density are not unique. This has been exploited to propose different  neutral densities that enjoy desirable properties in addition to near-neutrality, such as approximate conservation for adiabatic motion \citep[see][]{mcdo-jack97,desz-spri,tail16,kloc-et-al,stan19b,stan19,lang-et-al,stan21}.

The objectives of this paper are different. We aim to explore the consequences for ocean transport of imposing a constraint of the form \eqn{perp} on the velocity of water parcels. This constraint makes the transport locally two-dimensional \ch{in the sense that the velocity and thus infinitesimal displacements are restricted to (two-dimensional) planes. 
However, because of the non-integrability of the neutral planes, there are no neutral surfaces everywhere tangent to the neutral planes. Therefore, finite displacements are not confined to surfaces and parcel trajectories can explore the ocean in its full three dimensionality. In a topological sense, the transport is therefore globally three dimensional. The neutral planes are not integrable, but almost so. This results in a form of three-dimensional transport that is highly anisotropic and somewhat  unfamiliar.} 

The specific choice \eqn{neutralvec} for the dianeutral vector $\bm{n}$ is heuristic and controversial. \citet{nyca11} shows that the corresponding neutral displacements require mechanical forcing whereas displacements satisfying a similar but different condition $\bm{n}_P \cdot \bu=0$, with $\bm{n}_P$ defined by an analogue of \eqn{Mcplanes} involving enthalpy instead of density, do not \citep[see also][]{tail-wolf}. 
%This suggests that $\bm{n}_P$ may be a better choice of dianeutral vector than $\bm{n}$.  
Since the mathematical structure of the problem is the same whether \ch{$\bm{n}$, $\bm{n}_P$} or other alternatives are used, the distinction does not affect most of the arguments in this paper.  We therefore refer to motion with velocity constrained to planes, however chosen, as neutral transport, \ch{and we consistently denote the dianeutral vector by $\bm{n}$, regardless of its specific definition}. For the numerical experiments we report, we use \citeauthor{mcdo87}'s definition \eqn{neutralvec}   because it is the most widely adopted.

Neutral transport is a conceptual rather than literal model of ocean transport. It ignores many physical effects. It has a contradictory relation with the notion of adiabatic transport: adiabaticity is central to the local argument leading to the definition of neutral planes but it is not respected over finite-length trajectories. In appendix \ref{app:interpr}, we discuss a \ch{heuristic} interpretation of neutral trajectories as resulting from the continuous relaxation of the potential temperature and salinity of parcels to those of their surroundings. This interpretation supports the view that particle trajectories can usefully be thought of as neutral.   

The aims of this paper are twofold. The first aim is to connect neutral transport to a wealth of relevant mathematical results. The non-integrable field of planes $\bm{n}$ and the restriction to paths that are everywhere tangent to these planes equip the ocean, viewed as a subset of $\mathbb{R}^3$, with a contact structure \citep[e.g.][]{arno89,thur97}. Because the planes inherit a scalar product from the ambient Euclidean space, the ocean is also a sub-Riemannian manifold \citep[e.g.][]{stri86,mont02,cali-chan}. 

Contact geometry and sub-Riemannian geometry, here in their simplest, three-dimensional set up, provide the natural language to describe non-integrability and its consequences. We cast neutral transport in this language in \S\ref{sec:geometry}. There we detail the consequences for neutral transport of standard geometry results. These include the interpretation of non-integrability in terms of commutators of vector fields, the quantitative meaning of the helicity $\bm{n} \cdot \nabla \times \bm{n}$, and the implications of the Chow--Rashevskii accessibility theorem. We also discuss the natural notion of distance in sub-Riemannian manifolds -- the Carnot--Carath\'eodory (CC) distance -- and compute associated geodesics to illustrate some unfamiliar properties. The vast difference between CC and Euclidean distances separating points at different depths quantifies the anisotropy of  neutral transport. 
%The large values expected for the CC depth reflect how the near-integrability of the neutral planes restricts vertical transport in the ocean.

The second aim of this paper is to propose a  simple stochastic model of neutral transport. This model represents the motion of water parcels as a Brownian motion on the sub-Riemannian manifold defined by $\bm{n}$. The model is introduced and analysed in \S\ref{sec:stochastic}. It has a single parameter, the diffusivity $\kappa$ of the motion along the neutral planes. The corresponding diffusivity tensor is singular, with the span of $\bm{n}$ as its null space. As a result, the diffusion process is not strongly elliptic. The non-integrability of $\bm{n}$, equivalent to \citeauthor{horm67}'s (\citeyear{horm67}) condition, ensures that the process is instead hypoelliptic \citep[e.g.][]{jans-roge,pavl14}. This is also the case for the (slightly different) rotated diffusion that is widely used in ocean circulation models \citep{redi82,grif-et-al98} when the dianeutral diffusivity is set to zero and $\bm{n}$ is not integrable. 
A consequence of hypoellipticity is that water parcels explore the ocean in its full three-dimensionality. This is in sharp contrast with the integrable case, where parcels remain confined to (exactly) neutral surfaces.

We simulate the stochastic model numerically. The results highlight that, for short times, the dianeutral dispersion is qualitatively different from that obtained for a strongly anisotropic but non-singular diffusivity tensor. Long-time simulations show how parcels initialised at depth travel vast horizontal distances before reaching the mixed layer.  The numerical results are mainly intended as illustration. Large ensemble simulations would be required to reliably estimate quantities of oceanographic interest such as distributions of time and location of first passage to the surface. We leave this for future work.

\section{Geometry of neutral planes} \label{sec:geometry} 

\subsection{Notation}

It is natural to define the neutral planes using a field of $1$-forms (that is, covectors) rather than the vector field $\bm{n}$. This is how contact and sub-Riemannian geometry is usually formulated, to avoid the need for an unnecessary inner product in the  ambient space, $\mathbb{R}^3$ in our case. We therefore introduce the 
(Pfaffian) 1-form $\boeta$ dual to $\bm{n}$ (meaning that the pairing $\boeta(\bv) = \bm{n} \cdot \bv$ for all vectors $\bv$) and rewrite \eqn{perp} as
\beq
\boeta(\bu) = 0.
\lab{planes}
\eeq 
The choice \eqn{neutralvec} for $\bm{n}$ translates into
\begin{subequations} \lab{pfaff}
\begin{align}
\boeta &\propto R_\theta \, d \theta + R_S \, d S \\
&\propto R_\theta \, \partial_i \theta \, d x^i + R_S \, \partial_i S \, d x^i.
\lab{Mcplanes}
\end{align}
\end{subequations}
Here and \ch{in} what follows, summation over repeated indices is implied. In Eq.\ \ch{\eqn{pfaff}} we  abuse notation slightly by treating $\boeta$ indifferently as a 1-form in thermodynamic space $(\theta,S,p)$ and physical space $(x^1,x^2,x^3)$ when the latter is really the pull-back of the former by the map $(x^1,x^2,x^3) \mapsto (\theta,S,p)$.

We use a notation that accommodates both Cartesian and (geographic) spherical coordinates. We write $\bx=(x^1,x^2,x^3)$,
\beq
\ch{\bm{n}} = n^1(\bx) \, \bm{e}_1 + n^2(\bx) \, \bm{e}_2 + n^3(\bx) \, \bm{e}_3,
\eeq
and 
\beq
\ch{\boeta} = \eta_1(\bx) \, \bsigma^1 + \eta_2(\bx) \, \bsigma^2 + \eta_3(\bx) \, \bsigma^3,
\eeq
where  $\{\bm{e}_i\}$ is an orthonormal basis and $\{\bsigma^i\}$ is its dual basis, with $\bsigma^i(\bm{e}_j)=\delta^i_j$\ch{, where $\delta^i_j$ is the Kronecker delta}. With these bases, the components of $\bm{n}$ and $\boeta$ are identical:
\beq
\eta_i = n^i.
\eeq
In Cartesian coordinates, $(x^1,x^2,x^3)=(x,y,z)$,
\beq
(\bm{e}_1, \bm{e}_2, \bm{e}_3)  = ( \partial_x,\partial_y,\partial_z) \inter{and} (\bsigma^1, \bsigma^2, \bsigma^3) = (dx, dy, dz).
\eeq
In (geographic) spherical coordinates, $(x^1,x^2,x^3)=(\phi,\theta,z)$ are the longitude, latitude and altitude, 
\beq
(\bm{e}_1, \bm{e}_2, \bm{e}_3)  = ((a \cos \theta)^{-1} \partial_\phi, a^{-1} \partial_\theta, \partial_z) \inter{and} (\bsigma^1, \bsigma^2, \bsigma^3) = (a \cos \theta \, d \phi,  a \, d \theta, dz),
\eeq
where $a$ is the earth's radius. 

\ch{In what follows, we make use of both the vector $\bm{n}$ and its dual 1-form $\boeta$ to represent dianeutral direction and neutral planes. We use the `musical' notation 
\beq
\boeta = \bm{n}_\flat \quad \textrm{and} \quad \bm{n} = \boeta_\sharp
\eeq
to express the duality relation between $\bm{n}$ and $\boeta$. Details about this and other standard notation of differential geometry used in this paper can be found in \citet{schu80}, \cite{fran04} or, for an introduction centred on fluid dynamics, \citet{gilb-v18,gilb-v25} .}

\citet{mcdo87} and followers often formulate their results in terms of a `neutral gradient' $\nabla_n$ (or `projected non-orthogonal gradient'). This is best understood as a lifted differential $d_n$ such that, for any vector $\bm{v}$ in the $(x^1,x^2)$-plane and any function $f(\bx)$,
\beq
(d_n f)(\bm{v}) = (d f)(\bm{v}_\mathrm{lift}),
\eeq 
where $\bm{v}_\mathrm{lift} = \bm{v} - \boeta(\bm{v}) \bm{e}_3 / n^3$ is the neutral vector projecting onto $\bm{v}$. Explicitly, $d_n f$ is the 1-form on $\mathbb{R}^2$
\begin{subequations}
\begin{align}
d_n f &= d f - \partial_3 f \boeta / n^3 \\
&= (\partial_1 f- n^1 \partial_3 f/n^3) \, d x^1 + (\partial_2 f - n^2 \partial_3 f/n^3) \, d x^2.
\end{align}
\end{subequations}
With this definition,
\beq
R_\theta \, d_n \theta + R_S \, d_n S =  R_\theta \, d \theta + R_S \, d S - \frac{R_\theta \theta_z + R_S  S_z}{n^3} \boeta \ch{=0},
\eeq
\ch{since, from \eqn{pfaff}, $\boeta / n^3 = (R_\theta \, d \theta + R_S \, d S)/(R_\theta \theta_z + R_S S_z)$.
The} vanishing of the left-hand side can be taken as an implicit definition of $\boeta$ and the neutral planes.

% While \eqn{Mcplanes} is the most widely used definition of neutral planes, alternatives have been proposed.  \citet{nyca11} shows that displacements satisfying \eqn{displ} do require mechanical forcing whereas displacements satisfying a similar but different condition $\boeta_P(\bu)=0$, with $\boeta_P$ defined by an analogue of \eqn{Mcplanes} involving enthalpy instead of density, do not. Hence the velocity of water parcels is more plausibly constrained to the null space of $\boeta_P$ than to that of $\boeta$. 

\subsection{Non-integrability} \label{sec:noninte}

Condition \eqn{planes} determines what is called a distribution of planes, $\Delta$ say. A natural question is whether there exists a one-parameter family of surfaces everywhere tangent to these planes. This is the question of Frobenius integrability of the distribution \citep[e.g.][]{schu80,fran04}. A positive answer would imply the existence of neutral surfaces along which parcel motion is confined, so that neutral transport is truly two dimensional. A neutral density variable could then be defined in such a way that its level surfaces are the neutral surfaces. 

However, as pointed out by \citet{mcdo-jack88}, the distribution of neutral planes is not integrable: there are no surfaces everywhere tangent to the neutral planes \citep[see also][]{mcdo-jack97}. Non-integrability is generic: according to the Frobenius  theorem \citep{schu80,fran04}, integrability  requires
\beq
\boeta \wedge d \boeta = 0
\lab{integrability1}
\eeq
\ch{with $\wedge$ denoting the wedge product and $d$ the exterior derivative.}
The left-hand side is a 3-form which can be represented by a scalar, the helicity $\mathcal{H}$, defined by
\beq
\boeta \wedge d \boeta =  \mathcal{H}  \, \mu,
\lab{helicity1}
\eeq
where 
\beq
\mu = d x \wedge d y \wedge d z = \bsigma^1 \wedge \bsigma^2 \wedge \bsigma^3
\eeq
is the volume form.
This can also be written as 
\beq
\mathcal{H} = * (\boeta \wedge d \boeta) =  (*d\boeta)(\bm{n}),
\lab{helicity2}
\eeq 
where $*$ denotes the Hodge star operator \citep[e.g.][]{fran04}.
Integrability requires $\mathcal{H}=0$. 

For the Pfaffian \eqn{pfaff}, we have
\begin{subequations}
\begin{align}
\boeta \wedge d \boeta &\propto (R_S R_{\theta p} - R_\theta R_{Sp}) \, d \theta \wedge d S \wedge d p \\
&\propto R_S^2 (R_\theta/R_S)_p \, d \theta \wedge d S \wedge d p, \\
&\propto R_S^2 (R_\theta/R_S)_p \left| \frac{\partial(\theta,S,p)}{\partial(x^1,x^2,x^3)} \right| \, d x^1 \wedge d x^2 \wedge d x^3.
\end{align}
\end{subequations}
Hence the helicity satisfies
\beq
\mathcal{H} \propto R_S^2 (R_\theta/R_S)_p \left| \frac{\partial(\theta,S,p)}{\partial(x^1,x^2,x^3)} \right|.
\eeq
This does not vanish for a realistic equation of state for which $R_\theta/R_S = - \alpha/\beta$ is a function of $p$. 
However, as discussed by \citet{mcdo-jack07}, $\mathcal{H}$  is very small (in a suitable sense, see below) because the determinant $|\partial(\theta,S,p)/\partial(x^1,x^2,x^3)|$ turns out to be small in the ocean. Thus, while there are no strictly neutral surfaces, it is possible to construct surfaces that are approximately neutral in the sense that their tangent planes are almost neutral -- in other words, their normal vectors make a very small angle with $\bm{n}$. 
Much of the literature on neutral transport is devoted to the construction of such approximately neutral surfaces and the exploitation of the associated density variable \citep[e.g.][]{mcdo-jack97,desz-spri,tail16,kloc-et-al,stan19b,stan19,lang-et-al,stan21}.

\subsection{Contact structure}

\begin{figure}
\begin{center}
\includegraphics[height=6.5cm]{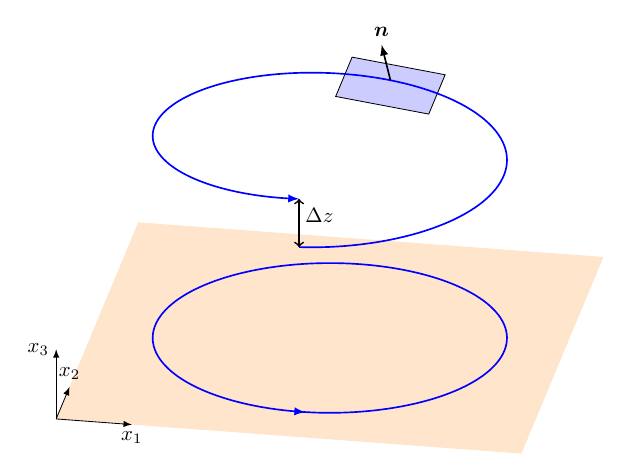}
\caption{Lift of a closed path in the $(x^1,x^2)$-plane to an open neutral trajectory.}
\label{fig:lift}
\end{center}
\end{figure}

The distribution of neutral planes defined by $\boeta$ equips the subset of $\mathbb{R}^3$ making up the ocean with a contact structure \citep[e.g.][]{arno89,thur97}. In the language of contact geometry, this makes the ocean a contact manifold and neutral trajectories  Legendrian submanifolds. 

Such trajectories $\bx(t)$ can be defined by lifting paths from the $(x^1,x^2)$-plane to $\mathbb{R}^3$, starting from some $\bx_0 = (x^1_0,x^2_0,z_0)$ and satisfying the constraint \eqn{planes} for $\bu = \dot{\bx}$. The non-integrability of the neutral planes implies that the lift of a path that is closed in the $(x^1,x^2)$-plane is not closed in $\mathbb{R}^3$. If we close the path by a vertical segment $(x^1_0,x^2_0,x^3 = z)$ with $z \in [z_0,z_1]$, we have that
\beq
\oint_{\partial S} \boeta = \int_{z_1}^{z_0} \boeta = \int_{z_1}^{z_0} n^3 \, d z = \int_S d \boeta,
\eeq
where $S$ is a surface bounded by the closed path and we use Stokes' theorem. See figure \ref{fig:lift} for an illustration. We can normalise $\boeta$ so that $n^3 = - 1$ to obtain %\matt{$z_0-z_1$ ? or $n^3 = -1$ ?}
\beq
\Delta z = z_1 - z_0 = \int_S d \boeta.
\eeq
This can be expressed in terms of the helicity by taking $S$ to be approximately neutral so that its normal is approximately $\bm{n}$. Parameterising $S$ as $(x^1,x^2,Z(x^1,x^2))$ for $(x^1,x^2) \in A$, we have
\begin{align}
\int_S d \boeta &= \int_S (* d \boeta)_\sharp \ip \mu \approx \int_{A} *d \boeta ( \bm{n})(x^1,x^2,Z(x^1,x^2)) \, d A \nonumber \\
& \approx  \int_{A} \mathcal{H}(x^1,x^2,Z(x^1,x^2)) \, d A, \lab{vertdispl}
\end{align}
where $dA$ is the area element in the $(x^1,x^2)$-plane \ch{and $\ip$ denotes the interior product, so that $(* d \boeta)_\sharp \ip \mu$ is the 2-form obtained by contracting $\mu$ with the vector $(* d \boeta)_\sharp$.
To obtain \eqn{vertdispl} we use} \eqn{helicity2} and the  parameterised form $\int_S \bm{v} \ip \mu = \int_A \bm{v} \cdot \bm{n} \, d A$ of surface integrals of vector fields.
%
%\beq
%\iint \left(\partial_1 \eta_2 - \partial_2 \eta_{1} - \partial_1 Z (\partial_2 \eta_{3} -\partial_3 \eta_{2}) - \partial_2 Z (\partial_1 \eta_{2}- \partial_2 \eta_{1})  \right) \, d x^1 \wedge  dx^2
%\eeq
%Since the surface $S$ can be taken close to neutral, 
%\beq
%(-Z_x,-Z_y,1) \, d x^1 \wedge d x^2 \approx (n^1,n^2,n^3) \, d S, 
%\eeq 
%where $dS = \bsigma^1 \wedge \bsigma^2$ is the area element in $(x^1,x^2)$,
%leading to the approximation
%\begin{subequations}
%\begin{align}
%z_1 - z_0 &\approx \iint \left( n^1 (n^{3y}-n^{2z}) + n^2 (n^{2x}-n^{1y}) + n^3 (n^{2x}-n^{1y}) \right) \, d S  \\
%&= \iint \mathcal{H}(x,y,Z(x,y)) \, d S \qquad  \textrm{for} \ n^3 = 1,
%\lab{vertdispl}
%\end{align}
%\end{subequations}
%where the last equality can be verified from the definition of $\mathcal{H}$ (or from the identity $\boeta \wedge d \boeta = (* d \boeta)(\bm{n}) \, \bsigma^1 \wedge \bsigma^2  \wedge \bsigma^3 )$. 
Results of this type, quantifying the gap in neutral trajectories opened by non-integrability, have been derived by \citet{mcdo-jack88}.

\subsection{Neutral vector fields}

\begin{figure}
\begin{center}
\includegraphics[height=3.5cm]{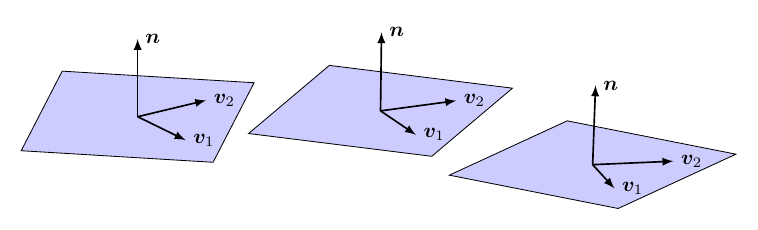}
\caption{The distribution $\Delta$ of neutral planes can be characterised by two vector fields $\bv_1$ and $\bv_2$.}
\label{fig:vecfields}
\end{center}
\end{figure}

The neutral planes are defined as the null space of the 1-form $\boeta$.  Alternatively, they can be characterised as the span of two linearly independent vector fields, $\bv_1$ and $\bv_2$ say, such that 
\beq
\boeta(\bv_1)=\boeta(\bv_2) = 0,
\lab{tanvec}
\eeq
see figure \ref{fig:vecfields}.
This alternative characterisation, standard in contact geometry, leads to useful insights, mainly through the formula 
\begin{subequations} \lab{detalie}
\begin{align}
d \boeta(\bv_1,\bv_2) &= \bv_1( \boeta(\bv_2)) - \bv_2( \boeta(\bv_1)) - \boeta([\bv_1,\bv_2]) \\
&= - \boeta([\bv_1,\bv_2]), \lab{detalieb}
\end{align}
\end{subequations}
where, in the first two terms, the vectors fields $\bv_i$ act on the scalars $\boeta(\bv_j)$ as derivatives, and $[\cdot,\cdot]$ is the Lie bracket, with coordinate expression
\beq
[\bv_1,\bv_2]^i = v^j_1 \partial_j v_2^i  - v^j_2 \partial_j v_1^i
\lab{commute}
\eeq
\citep[see e.g.][]{fran04}.

The condition $\boeta \wedge d \boeta = 0$ for integrability is equivalent to $d\boeta(\bv_1,\bv_2) = 0$ since, for any vector $\bv_3$, 
\begin{align}
(\boeta \wedge d \boeta)(\bv_1,\bv_2,\bv_3) &= \boeta(\bv_1) d \boeta(\bv_2,\bv_3) + \boeta(\bv_2) d \boeta(\bv_3,\bv_1) + \boeta(\bv_3) d \boeta(\bv_1,\bv_2) \nonumber \\
&= \boeta(\bv_3) d \boeta(\bv_1,\bv_2).
\lab{aa}
\end{align}
The identity \eqn{detalieb} then gives an alternative formulation of the integrability condition $\mathcal{H}=0$, namely
\beq
\boeta([\bv_1,\bv_2]) = 0 \quad \textrm{for all} \  \bv_1 \  \textrm{and} \ \bv_2 \  \textrm{satisfying} \ \eqn{tanvec}.
\lab{integrability2}
\eeq
This condition depends on $\bv_i$ at the point where it is evaluated, not on its derivatives, as \eqn{detalie} shows. 
\ch{In coordinates, 
\beq
\boeta([\bv_1,\bv_2]) = \eta_i (v_1^j \partial_j v_2^i - v_2^j \partial_j v_1^i) = v_1^i v_2^j (\partial_j \eta_i - \partial_i \eta_j) \lab{tanvec2},
\eeq}
where the first equality uses the derivative of \eqn{tanvec}.

The converse non-integrability condition
\beq
\boeta([\bv_1,\bv_2]) \not= 0,
\lab{bracketgenerating}
\eeq 
equivalent to $\mathcal{H}\not=0$, is the so-called `bracket generating' condition which expresses that $\bv_1$, $\bv_2$ and $[\bv_1,\bv_2]$ span the whole of $\mathbb{R}^3$. It leads to the interpretation of non-integrability as resulting from the fact that the commutator of two vectors tangent to the neutral planes is not itself tangent to the neutral planes. 

\begin{figure}
\begin{center}
\includegraphics[height=6.5cm]{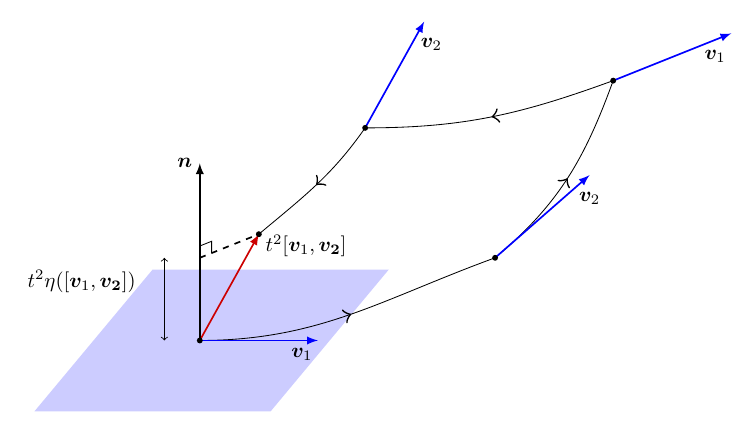}
\caption{Dianeutral transport, i.e.\ transport across the neutral plane (indicated by the blue shading), results from the successive transport along the vectors $\bv_1$, $\bv_2$ $-\bv_1$ and $-\bv_2$ for a short time $t \ll 1$. A parcel path is shown as the black line. Its endpoint, with position $t^2 [\bv_1,\bv_2] + O(t^3)$, has dianeutral component $t^2 \boeta([\bv_1,\bv_2]) + O(t^3)$.}
\label{fig:lie}
\end{center}
\end{figure}

In the oceanographic context, condition \eqn{bracketgenerating} is key to the existence of dianeutral transport, that is, transport perpendicular to the neutral planes, with a velocity field that is everywhere neutral.
Since the commutator \eqn{commute} can be thought of as the $O(t^2)$ net displacement that emerges for a trajectory flowing for a small time $t$ along each of $\bv_1$, $\bv_2$, $-\bv_1$ and $-\bv_2$ in succession \citep[e.g.][]{fran04}, dianeutral transport can be achieved by  neutral trajectories that flow successively along $\pm \bv_1$ and $\pm \bv_2$ in this manner. See figure \ref{fig:lie} for an illustration.

The helicity $\mathcal{H}$ provides a measure of non-integrability: we show below that
\beq
\mathcal{H} =  - \frac{\boeta([\bv_1,\bv_2])}{ \mu(\bv_1,\bv_2,\bm{n})},
\lab{helicityin}
\eeq
where  $\mu(\bv_1,\bv_2,\bm{n})$ is the (signed) volume of the parallelepiped spanned by $\bv_1, \, \bv_2$ and $\bm{n}$. Since $\bm{n} \cdot \bv_1 = \bm{n} \cdot \bv_2 = 0$, this volume can otherwise be interpreted as  the (signed) area spanned by the vectors $\bv_1$ and $\bv_2$ if $\bm{n}$ is normalised so that $|\bm{n}|^2=|\boeta|^2 = (n^1)^2 + (n^2)^2 + (n^3)^2 =1$.

% \begin{figure}
% \begin{center}
% \includegraphics[height=4cm]{nz.png}
% \caption{Vertical component $n^3$ of the vector field $\bm{n}$ characterising the neutral planes at $z=-1300$ m.}  
% \label{fig:nz}
% \end{center}
% \end{figure}

Eq.\ \eqn{helicityin} shows that $\mathcal{H}$ is the ratio of the  dianeutral displacement realised by an infinitesimal trajectory to the area of the projection of this trajectory on the neutral plane. 
This interpretation requires the normalisation $|\bm{n}|^2 =1$ whereas \eqn{vertdispl} requires $n^3 = - 1$. Vertical gradients of $\theta$ and $S$ in the ocean far exceed horizontal gradients so that $n^3 \gg n^1,\, n^2$ and the neutral planes are very close to horizontal. %, see figure \ref{fig:nz}. 
As a result, \ch{there is little difference between the two normalisations, and approximation \eqn{vertdispl} also applies with the normalisation $|\bm{n}|=1$ (for the sign choice $n^3 < 0$)}. 

\ch{Note that for dimensionless $\bm{n}$, as is the case with the normalisation $|\bm{n}|=1$, $\mathcal{H}$ has dimension of $(\textrm{length})^{-1}$. In contrast, if $\bm{n}$ is taken as $\bm{n} = (R_\theta \nabla \theta + R_S \nabla S)/R$, as is often the case \citep{mcdo87}, the dimension of $\bm{n} \cdot \nabla \times \bm{n}$ is   $(\textrm{length})^{-3}$.}

To establish \eqn{helicityin}, note that the normalisation implies that $\boeta(\bm{n})=1$ and therefore that
\beq
(\boeta \wedge \d \boeta)(\bv_1,\bv_2,\bm{n}) = \boeta(\bm{n}) d \boeta(\bv_1,\bv_2) = - \boeta([\bv_1,\bv_2]),
\eeq
on using \eqn{aa}.
It follows from this and \eqn{helicity1} that
\beq
\boeta([\bv_1,\bv_2]) =- (\boeta \wedge d \boeta)(\bv_1,\bv_2,\bm{n}) = - \mathcal{H} \, \mu(\bv_1,\bv_2,\bm{n}),
\eeq
yielding \eqn{helicityin}.  

We can characterise the magnitude of the helicity in a manner independent of the two vectors $\bv_1$ and $\bv_2$. 
\ch{Observe first that  helicity is independent of the particular choice of vectors $\bm{v}_1$ and $\bm{v}_2$ spanning the neutral plane. We can therefore pick these vectors as the unit vectors that maximise $|\boeta([\bv_1,\bv_2])| = |\mathcal{H} \mu(\bv_1,\bv_2,\bm{n})|$. This gives
\beq
|\mathcal{H}| = \max_{\substack{|\bv_1|=|\bv_2|=1 \\ \boeta(\bv_1)=\boeta(\bv_2) = 0}} |\boeta([\bv_1,\bv_2])|,
\eeq
on noting that $|\mu(\bv_1,\bv_2,\bm{n})|=1$ when maximised.} 
%\textcolor{purple}{Should we briefly say that if $|\mu(\bv_1,\bv_2,\bm{n})|\neq1$, then there exist reals $\alpha, \beta$ and a unite vector $v_2’$ such that $ v_2 = \alpha v_2’ + \beta v_1$. Since $|v_2| = 1, \ \alpha^2 + \beta^2 = 1$ and $|\eta\lp \left[ v_1, v_2\right]\rp| = |\alpha| |\eta\lp \left[ v_1, v_2’\right]\rp|$, with $|\alpha|<1$ meaning that $|\eta\lp \left[ v_1, v_2\right]\rp|< |\eta\lp \left[ v_1, v_2’\right]\rp|  $ : $|\eta\lp \left[ v_1, v_2\right]\rp| $ is therefore not maximised. }

\begin{figure}
\begin{center}
\includegraphics[scale = 1
]{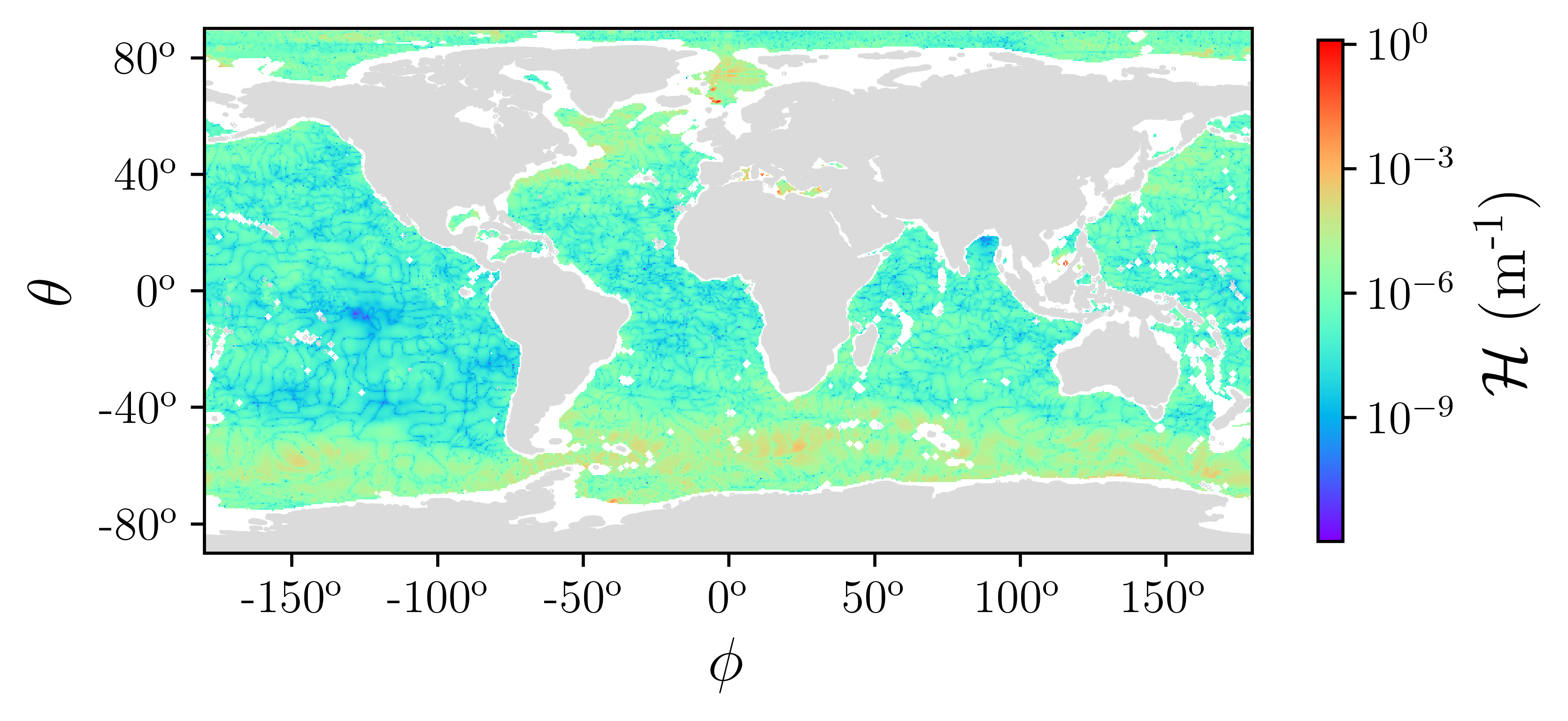}
\caption{\ch{Magnitude $|\mathcal{H}|$ of helicity} at $z=-1300$ m computed from WOCE data.}
\label{fig:Helicity}
\end{center}
\end{figure}

We compute the ocean helicity $\mathcal{H}$ starting from a gridded representation of $\bm{n}$ derived from the WOCE climatology
\citep{gour-kolt}.  We obtain  a  spherical coordinate expression for $\mathcal{H}$ directly from \eqn{helicity1}. Taking the exterior derivative of
\beq
\boeta = a \cos \theta \, n^1 \,  d\phi + a \, n^2 \, d \theta + n^3 \, d z,
\eeq
we  find
\begin{align}
d \boeta &= a ( \partial_\phi n^2 - \cos \theta \, \partial_\theta n^1 + \sin \theta \, n^1) \, d \phi \wedge d \theta + (\partial_\theta n^3 - a \partial_z n^2) \, d \theta \wedge d z \nonumber \\
&+ (a \cos \theta \, \partial_z n^1 - \partial_\phi n^3) \, d z \wedge d \phi.
\end{align}
Applying $\boeta \, \wedge$ to this and using that 
\beq
\mu = a^2 \cos \theta \, d \phi  \wedge d \theta \wedge dz
\eeq
we obtain
\begin{align}
\mathcal{H} &= n^1 \left( \frac{1}{a} \partial_\theta n^3  - \partial_z n^2 \right) 
+ n^2 \left(\partial_z n^1 - \frac{1}{a \cos \theta} \partial_\phi n^3 \right) + n^3 \left(  \frac{1}{a \cos \theta} \partial_\phi n^2 - \frac{1}{a} \partial_\theta n^1 + \tan \theta \, n^1 \right) \nonumber \\
&= \frac{1}{a \cos \theta} \left( n^3 \partial_\phi n^2 - n^2 \partial_\phi n^3 \right) +
\frac{1}{a} \left( n^1 \partial_\theta n^3 - \frac{n^3}{\cos \theta} \partial_\theta (\cos \theta \, n^1) \right) 
+  n^2 \partial_z n^1 - n^2 \partial_z n^1. 
%\\ & \ +\matt{\frac{1}{a} \lp n^1\partial_\theta n^3 - \frac{n^3}{\cos\theta} \partial_\theta\lp \cos\theta n^1\rp\rp + n^2\partial_z n^1-n^1\partial_z n^2}
\end{align}
\ch{Values of $\mathcal{H}$ on the same grid as those of $\bm{n}$ are computed using centred finite differences.}

Figure \ref{fig:Helicity} shows the value of $\mathcal{H}$ at $z=-1300$ m. Typical values at this depth are in the range $(10^{-7},10^{-5})$ m$^{-1}$, with much larger values in the Southern Ocean. \ch{A value of $\mathcal{H} = 10^{-6}$ m$^{-1}$, for instance, implies that a vertical gap of  $1$ m opens for a neutral trajectory that projects onto an area of $1$ km$^2$. Trajectories that enclose much larger areas can in principle lead to more significant vertical displacements according to \eqn{vertdispl}, but one must then account for the multiple sign changes in $\mathcal{H}$, leading to large cancellations. These sign changes are visible in figure \ref{fig:Helicity}: the network of dark blue contours indicates where $\mathcal{H} \approx 0$ and, overall, $\mathcal{H} > 0$ in 50.2 \% of area and $\mathcal{H} < 0$ in the rest.
 As a result of cancellations, basin-scale closed trajectories lead to gaps of tens to hundreds of meters only \citep{mcdo-jack88}. This exemplifies the weakness of dianeutral transport in the ocean that results from the near-integrability of the neutral planes.  
We illustrate this further in \S\ref{sec:stochastic} by computing long (random) neutral paths.}

\ch{We note that $\mathcal{H}$ depends on second spatial derivatives of the temperature and salinity fields and so is sensitive to the choice of climatological dataset and to numerical details. We therefore regard conclusions drawn on the basis of the values of $\mathcal{H}$ displayed in figure \ref{fig:Helicity}  as qualitative rather than quantitative.}

\subsection{Accessibility}

The non-integrability, expressed as $[\bv_1,\bv_2] \not= 0$, implies that $\bv_1, \, \bv_2$ and $[\bv_1,\bv_2]$ span the whole of $\mathbb{R}^3$. A consequence, embodied in the accessibility theorem of \citet{cara09} generalised by \citet{chow39} and \citet{rash38}, is that any two points in $\mathbb{R}^3$ can be joined by paths that are everywhere tangent to the neutral planes. In other words, any pair of points in the ocean can be joined by at least one (in practice, infinitely many) neutral trajectories.  In this topological sense, neutral motion is fully three-dimensional.
This was pointed out by \citet{benn19}. 

\subsection{Sub-Riemannian geometry} 

While the condition of neutrality does not restrict the topology of ocean transport, it severely restricts its geometry. 
To examine this requires to consider the ocean as both a contact manifold and a sub-Riemannian manifold, that is, to consider a metric defined for neutral vectors. Because $\mathbb{R}^3$ is equipped with the usual Euclidean metric (already used in the interpretation of the helicity), the distribution of neutral planes simply inherits this metric. 

The sub-Riemannian geometry of a manifold is encoded in the co-metric, a twice contravariant tensor $h$ that maps 1-forms to vectors tangent to the neutral planes, that is,  $h: T_*M \to \Delta$, according to
\beq
h(\bm{\alpha})\cdot \bv = \bm{\alpha}(\bv) \quad \textrm{for all} \ \ \bv \in \Delta.
\eeq
An explicit form for $h$ is obtained using two  neutral vector fields $\bv_1$  and $\bv_2$ that serve as a basis for vectors in $\Delta$. It is given by
\beq
h = h^{ij} \bm{e}_i \otimes \bm{e}_j = g^{rs} \bv_r \otimes \bv_s = g^{rs} v^i_{r} v^j_{s} \, \bm{e}_i \otimes \bm{e}_j,
\eeq 
where $g^{rs}$ is the inverse of the Gram matrix $g_{rs} = \bv_{r} \cdot \bv_s$.
To check this, note that
\beq
h(\bm{\alpha}) =  g^{rs} \bm{\alpha}(\bv_s) \bv_r,
\eeq
since this implies
\beq
h(\bm{\alpha}) \cdot \bv = g^{rs} \bm{\alpha}(\bv_s) (\bv_r \cdot \bv) = \bm{\alpha} \left(g^{rs} (\bv_r \cdot \bv) \bv_s \right) = \bm{\alpha}(\bv),
\eeq
using that the expansion of $\bv$ in the basis $(\bv_1,\bv_2)$ is $\bv = g^{rs} (\bv_r \cdot \bv) \bv_s$.

We can take $\bv_1 = \bm{e}_1 \times \bm{n} = (0,-n^3,n^2)^\intercal$ and $\bv_2 =\bm{e}_2 \times \bm{n} = (n^3,0,-n^1)^\intercal$ with $|\bm{n}|=1$ and find (in the basis $(\bm{e}_1,\bm{e}_2,\bm{e}_3)$)
\beq
h =  \mathbb{I} - \bm{n} \otimes \bm{n} =  \begin{pmatrix} 
(n^2)^2 + (n^3)^2 & - n^1 n^2 & - n^1 n^3 \\
-n^1 n^2 & (n^1)^2 + (n^3)^2 & - n^2 n^3 \\
-n^1 n^3 & -n^2 n^3 & (n^1)^2 + (n^2)^2
\end{pmatrix},
\lab{hmatrix}
\eeq
\ch{on using that $1-(n^1)^2 = (n^2)^2 + (n^3)^2$ and similar.}
This is a rank-2 matrix. Its null space is spanned by $\boeta$ and its image is the null space of $\boeta$. 

The integrability conditions \eqn{integrability1} or \eqn{integrability2} can be rephrased as a degeneracy condition on the raised Christoffel symbols 
\beq
\Gamma^{ijk} = \frac{1}{2} \left( h^{lj} \partial_l {h^{ik}} + h^{lk} \partial_l{h^{ij}} - h^{li} \partial_l {h^{jk}}  \right),
\eeq
associated with $h$ \citep{stri86}.\footnote{In a correction, \citet{stri89} defines the symbols with the opposite sign. We follow the convention of \citet{cali-chan}.}
See Appendix \ref{app:chrisoffel} for details. 

\subsection{Carnot--Carath\'eodory distance and geodesics} \label{sec:CC}

Given that any two points in the ocean can be joined by a neutral path, it is natural to ask what the length of the shortest \ch{neutral} path is. 
This is the Carnot--Carath\'eodory (CC) distance 
\beq
\dcc(\bm{x},\bm{y}) = \inf_{\bm{\varphi}(\cdot)} \int_0^1 | \dot{\bm{\varphi}}(t)| \, \d t \quad \textrm{with} \ \ \bm{\varphi}(0)=\bx, \ \ \bm{\varphi}(1)=\by \ \ \textrm{and} \ \ \boeta(\dot{\bm{\varphi}}(t))=0.
\lab{dcc}
\eeq
Geodesics, that is, distance-minimising \ch{neutral} paths can be obtained by solving the Hamiltonian system
\beq
\dot{\bx} = \partial_{\bm{p}} H, \quad \dot{\bm{p}} = - \partial_{\bx} H, \quad \textrm{where} \ \ H(\bx,\bm{p}) = \tfrac{1}{2} h^{ij}(\bx) p_i p_j
\lab{hamilton}
\eeq
\cite[e.g.][]{cali-chan}.
All minimising paths can be found by solving these equations in the case of a contact structure (although it is not the case for general sub-Riemannian manifolds \citep{mont02}.) \ch{Eqs.\ \eqn{hamilton} can be solved as a two-point boundary-value problem prescribing $\bx(0)$ and $\bx(1)$, or as an initial-value problem prescribing $\bx(0)$ and $\bm{p}(0)$. We solve the latter problem below.} 

Sub-Riemannian geodesics are  unfamiliar in that, at each location, there is not just one geodesic departing in a given neutral direction but a one-parameter family of such geodesics. \ch{This arises because the initial momentum $\bm{p}(0)$, which determines the geodesic path starting at some $\bx(0)$, is not in one-to-one correspondence with the initial velocity. Indeed, from \eqn{hamilton}, $\dot{\bx}(0) = h(\bx(0)) \, \bm{p}(0)$, and $h$ is singular. This in contrast with the Riemannian case where momentum and velocity are equivalent.} 

\ch{The family of geodesics starting in a given direction is parameterised by the component of $\bm{p}(0)$ parallel to $\boeta$. This component has no impact on the initial direction of the geodesic since the right-hand side of $\dot{\bx}(0) = h(\bx(0)) \, \bm{p}(0)$ is  invariant under the transformation $\bm{p}(0) \mapsto \bm{p}(0) + k \boeta$ for any $k \in \mathbb{R}$. Geodesics corresponding to different values of this component depart from one other quadratically with $t$ for small $t$ (see Appendix \ref{app:local} for details). Note that the additional degree of freedom associated with the component of $\bm{p}(0)$ along $\boeta$ is crucial to ensure that any two points of the ocean can  be joined locally by a geodesic trajectory: a three-parameter family of paths from any point is required to reach any other point of the three-dimensional ambient space.
}
%Geodesics that differ only by the value of $k$ diverge but with a distance $\dcc$ between them that increases only as $t^{3/2}$ for small $t$ (rather than $t$ for geodesic starting in different directions).  

Over a small time $t \ll 1$, paths tangent to $\Delta$ travel distances proportional to $t$ in directions parallel to $\Delta$ but proportional to $\mathcal{H} t^2$ in the perpendicular direction (see Appendix \ref{app:local}). As a result, the CC distance is highly anisotropic,  equivalent to a distance of the form \ch{$|\bx_\parallel| + |x_\perp/\mathcal{H}|^{1/2}$, where $\bx_\parallel \in \mathbb{R}^2$ and $x_\perp \in \mathbb{R}$ are local coordinates along and perpendicular to $\Delta$ \citep{stri86,grom96,mont02}. In the oceanographic context, $x_\perp \approx z$.}

To gain  insight into the anisotropy of geodesics in the ocean, we compute a few geodesics using WOCE climatological data for $\boeta$.  We use the spherical coordinates $(\phi,\theta,z)$ and their conjugate $(p_\phi, p_\theta, p_z)$, with
\beq
\bm{p} = p_1 \, \bsigma_1 + p_2 \, \bsigma_2 + p_3 \bsigma_3 = p_\phi \, d \phi + p_\theta \, d \theta + p_z \, d z,
\eeq
and therefore
\beq
(p_\phi, p_\theta, p_z) = (a \cos \theta \, p_1, a \, p_2, p_3).
\eeq
In terms of these, the Hamiltonian is
\begin{subequations}
\begin{align}
H &= \frac{1}{2}  (p_\phi, p_\theta, p_z) \begin{pmatrix} 
\frac{(n^2)^2 + (n^3)^2}{a^2 \cos^2 \theta}  & - \frac{n^1 n^2}{a^2 \cos \theta} & - \frac{n^1 n^3}{a \cos \theta} \\
-\frac{n^1 n^2}{a^2 \cos \theta} & \frac{(n^1)^2 + (n^3)^2}{a^2} & - \frac{n^2 n^3}{a} \\
- \frac{n^1 n^3}{a \cos \theta} & -\frac{n^2 n^3}{a} & (n^1)^2 + (n^2)^2
\end{pmatrix} 
  \begin{pmatrix} p_\phi \\ p_\theta \\  p_z \end{pmatrix} \\
  &= \frac{1}{2} \left( \left(\frac{n^2}{a \cos \theta} p_\phi - \frac{n^1}{a} p_\theta \right)^2 + \left( \frac{n^3}{a} p_\theta - n^2 p_z \right)^2 + \left( n^1 p_z - \frac{n^3}{a \cos \theta} p_\phi \right)^2 \right).
\end{align}
\end{subequations}
The geodesic equations follow from \eqn{hamilton}.
% \beq
% \begin{pmatrix}
% \dot \phi \\ \dot \theta \\ \dot z
% \end{pmatrix} = \begin{pmatrix} 
% \frac{n^2^2 + n^3^2}{a^2 \cos^2 \theta}  & - \frac{n^1 n^2}{a^2 \cos \theta} & - \frac{n^1 n^3}{a \cos \theta} \\
% -\frac{n^1 n^2}{a^2 \cos \theta} & \frac{n^1^2 + n^3^2}{a^2} & - \frac{n^2 n^3}{a} \\
% - \frac{n^1 n^3}{a \cos \theta} & -\frac{n^2 n^3}{a} & n^1^2 + n^2^2
% \end{pmatrix} 
%   \begin{pmatrix} p_\phi \\ p_\theta \\  p_z \end{pmatrix}. 
% \eeq

\begin{figure}
    \centering
    \includegraphics[scale = 1]{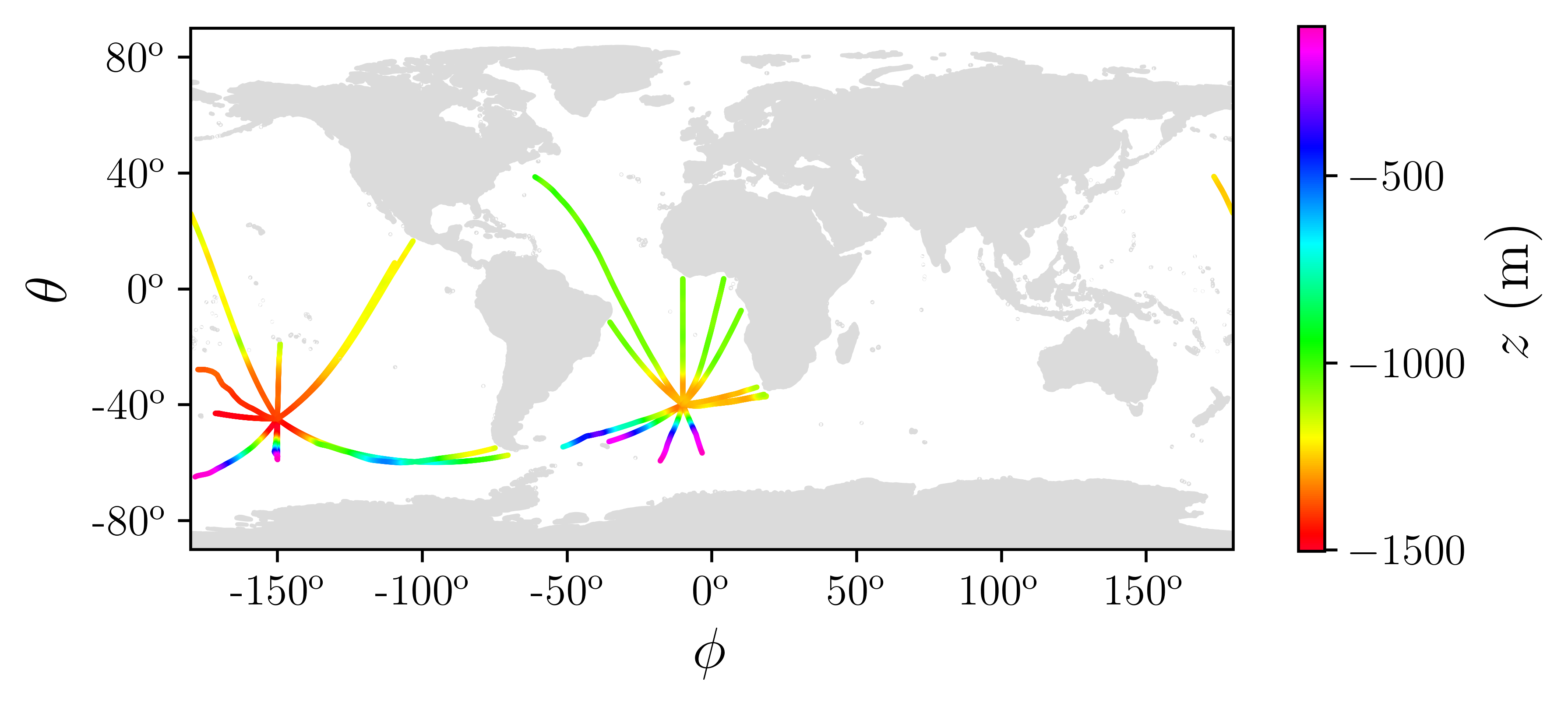}
    \caption{Neutral geodesics starting from a point in the Pacific and from a point in the Atlantic, both at a depth $z = - 1400$ m. The colour indicates the depth along each path.}  
    \label{fig:geodesics_example}
\end{figure}

We solve the geodesic equations using Heun's method (two-stage Runge--Kutta method) and a tri-linear interpolation to approximate the components of $\bm{n}$ away from grid points. We verify that the resulting trajectories are neutral by computing the dot product $\dot{\bm{x}} \cdot \bm{n}$ between the velocity and the dianeutral direction, checking that it vanishes up to numerical errors. Geodesics emanating from two points with $z=-1400$ m, one in the Atlantic and the other in the Pacific, are shown in figure \ref{fig:geodesics_example}. The figure illustrates how geodesics are near horizontal at mid-latitudes but much steeper in the Southern Ocean, primarily because of sloping  neutral surfaces  and, to a lesser extent, because of large values of $|\mathcal{H}|$. 

\begin{figure}
    \centering
    \includegraphics[scale = .8]{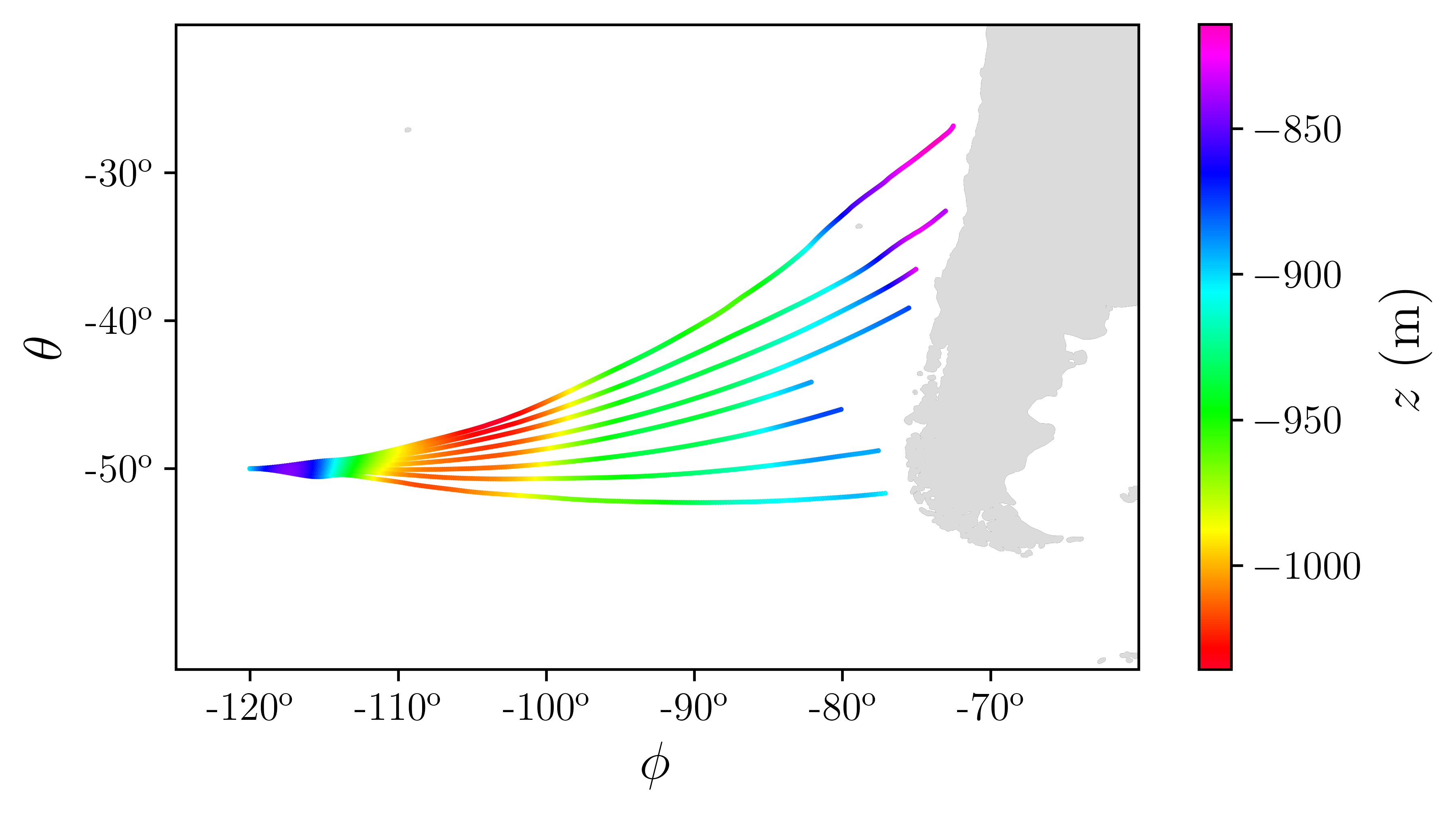}
    \caption{Neutral geodesics emanating from a single point in a given direction, differing by the component of their initial momentum $\bm{p}$ in the neutral direction $\boeta$.}
    \label{fig:family}
\end{figure}

As mentioned above, there is a one-parameter family of geodesics starting from any point in a given direction. Figure \ref{fig:geodesics_example} only displays a single arbitrary member of his family. To complement this, we show in figure \ref{fig:family} several members of the family of geodesics starting at $\phi=-120^\circ$, $\theta = -50^\circ$ and $z=-1400$ m. These are obtained by adding to the initial momentum $\bm{p}$ multiples of the local $\boeta$. \ch{The geodesics depart from each other quadratically with $t$ for small $t$, as expected,
and diverge quickly.}

% [[SHOW EXAMPLES OF GEODESICS, SAY SOMETHING ABOUT NUMERICS]]
% \matt{ To numerically integrate geodesic equations, we chose to use Heun’s method also know as a two stage Runge-Kutta method. Different methods have been used to verify the accuracy of numerical integration. One of them is by computing the scalar product of the particle velocity with the normal vector and verify it is close to zero to $10^{-10}$ .  }

An \ch{interesting} concept is what can be termed the Carnot--Carath\'eodory depth of the ocean, namely the minimum CC distance between a point at the bottom of the ocean and the surface. This depth, clearly much larger than the physical depth, quantifies the overall vertical transport that results from the non-integrability of the neutral planes (in the integrable case, it would be $\infty$ for non-outcropping neutral surfaces). We do not attempt to evaluate the CC depth as this would require tackling a formidable optimisation problem. Because most geodesic paths intersect the continental slope, it would also be necessary to determine a physically meaningful reflection law. 

% Here, based on the short geodesic segments we computed, we only suggest an order of magnitude of XXX km for the CC depth of the ocean. This estimate needs confirming by detailed computations.  

CC distances are the shortest possible distances between two points that water parcels can travel with a velocity that is always neutral. It is natural ask what the typical rather than shortest distances are. In the next section, we develop a simple stochastic model addressing this question.

\section{A stochastic model} \label{sec:stochastic}

\subsection{Formulation}

We construct a  stochastic model representing neutral motion, physically quasi two-dimensional turbulence, by a constant (neutral) diffusivity $\kappa$. The position $\bX(t)$ of a water parcel obeys the Stratonovich stochastic differential equation (SDE)
\beq
\d \bX = \sqrt{2\kappa} \, \bm{n}(\bX) \times \circ \, \d \bm{W},
\lab{smodel1}
\eeq
where $\bm{W} = (W_1, W_2,W_3)$ is a vector of independent Wiener processes and $\circ$ signals the Stratonovich interpretation \ch{\citep[e.g.][]{gard04,evan12,pavl14}}.
Expanding the cross product, this can also be written as
\beq
\d \bX = \sqrt{2 \kappa} \left( \bv_1(\bX) \circ \d W_1 + \bv_2(\bX) \circ \d W_2 + \bv_3(\bX) \circ \d W_3 \right),
\lab{smodel1+}
\eeq
where 
\beq
\bv_1 =  \bm{n}\times  \bm{e}_1  = \begin{pmatrix} 0 \\ n^3 \\ - n^2 \end{pmatrix}, \quad
\bv_2 =  \bm{n} \times \bm{e}_2  = \begin{pmatrix} -n^3 \\ 0 \\ n^1 \end{pmatrix}, \quad
\bv_3 =  \bm{n}  \times \bm{e}_3 = \begin{pmatrix} n^2 \\ -n^1 \\ 0 \end{pmatrix}
\lab{3vectors}
\eeq
are three neutral vector fields. 

An alternative to the cross-product construction in \eqn{smodel1} uses an orthogonal projection to construct the model
\beq
\d \bX = \sqrt{2\kappa} \, (\mathbb{I} - \bm{n}(\bX) \otimes \bm{n}(\bX)) \circ \d \bm{W},
\lab{smodel2}
\eeq
or in coordinates 
\beq
\d X^i = \sqrt{2\kappa} \, h^{ij}(\bX) \circ \d W_j,
\eeq
with the matrix $h^{ij}$ defined in \eqn{hmatrix}.
The two models \eqn{smodel1} and \eqn{smodel2} turn out to be equivalent, see Appendix \ref{app:generator}.
%: in both cases that increments are zero-mean Gaussian with covariance
%\beq
%\E \left( \d X^i \d X^j \right) = 2 \kappa \,  h^{ij}(\bX) \, \d t.
%\eeq 

It is essential that the SDEs \eqn{smodel1} and \eqn{smodel2} be interpreted in the sense of Stratonovich for the motion to be truly neutral. The inadequacy of an It\^o interpretation is illustrated by considering the case of an integrable distribution of neutral planes. This arises when
\beq
\boeta = d f / |\nabla f|
\lab{intnormal}
\eeq
for some function $f$. The level sets of $f$ are then (exactly) neutral surfaces. The trajectories $\bX(t)$ should be confined to these surfaces. They are with the Stratonovich interpretation,
\beq
\d f = \partial_i f(\bX) \circ \d X^i = \sqrt{2 \kappa} |\nabla f| \eta_i(\bX) \,  h^{ij}(\bX) \circ \, \d W_j  = 0,
\eeq
but not in the It\^o interpretation,
\begin{align}
\d f &= \sqrt{2 \kappa} |\nabla f|\,  \eta_i(\bX) \,  h^{ij}(\bX)  \, \d W_j + \kappa \partial_{ij} f(\bX) \,  h^{ij}(\bX) \, \d t \nonumber \\
& = \kappa \partial_{ij} f(\bX) \,h^{ij}(\bX) \, \d t \not= 0,
\end{align}
\ch{where we have used It\^o's formula \citep[e.g.][]{gard04,evan12,pavl14}.}
When \eqn{intnormal} holds, the models \eqn{smodel1} and \eqn{smodel2} describe Brownian motion on the surfaces $f = \mathrm{const}$, see
\citet{vand-lewi} and \citet[][\S V.31]{roge-will00}.

The It\^o equivalent to \eqn{smodel1} is 
\beq
\d \bX = - \kappa (\nabla \cdot \bm{n}) \bm{n} \, \d t + \sqrt{2\kappa} \, \bm{n}(\bX) \times  \d \bm{W}
\lab{smodel1Ito}
\eeq
(see Appendix \ref{app:generator}).

\subsection{Properties}

The process $\bX(t)$ is characterised by its generator $\mathcal{L}$ such that observables of the form $u(\bx,t) = \E_{\bx} \left[g(\bX(t))\right]$, where $g$ is an arbitrary function and $\E_{\bx}$ denotes expectation for trajectories starting at $\bX(0)=\bx$, satisfy
\beq
\partial_t u = \mathcal{L} u.
\eeq
The transition probability $p(\bx,t|\by)$ correspondingly satisfies
%\textcolor{red}{is this notation ok? not confusing with $\mathbf{p}$? alternative: $\mathfrak{p} \quad$}
\beq
\partial_t p = \mathcal{L}^* p,
\eeq
where $\mathcal{L}^*$ is the adjoint of $\mathcal{L}$. For a Stratonovich equation of the form \eqn{smodel1+}, the generator is given by 
\beq
\mathcal{L} = \kappa \left( (\bv_1 \cdot \nabla)^2 + (\bv_2 \cdot \nabla)^2 + (\bv_3 \cdot \nabla)^2 \right)
\eeq
\citep[e.g.][]{roge-will00,pavl14}.
This can be written explicitly as 
\beq
\mathcal{L} = \kappa \left( h^{ij} \partial_{ij} - (\nabla \cdot \bm{n}) n^j \partial_j \right)
\eeq
on using the form \eqn{3vectors} of the vector fields $\bv_k$, see Appendix \ref{app:generator}.

Because $h^{ij}$ is singular, the operator $\mathcal{L}$ is not  elliptic and the diffusion process it represents is degenerate. However, when $\mathcal{H} \not = 0$, $\mathcal{L}$ satisfies the bracket generating condition \eqn{bracketgenerating} which \citet{horm67}
showed to imply hypoellipticity, guaranteeing a smooth transition probability $p(\bx,t|\by)$ (see, e.g., \citet[][\S V.38]{roge-will00} or \citet[][\S 6.1.2]{pavl14}. 
Physically, this means that water parcels released at a point $\by$ of the ocean and moving according to \eqn{smodel1} have a non-zero probability of reaching any other point $\bx$ some time later. Thus parcels explore the ocean in its full three dimensionality, in spite of their velocity being constrained to the two-dimensional neutral at all times. 

It is standard  in  numerical models of the ocean to parametrise the mixing of scalars as a diffusion represented by the self-adjoint generator
\beq
\tilde{\mathcal{L}} = \kappa \partial_i \left( h^{ij}  \partial_j\right) = \kappa \partial_i ((\delta_{ij} - n^i n^j) \partial_j)
\lab{redi}
\eeq
and different choices of $\bm{n}$, \ch{e.g.\  perpendicular to isopycnals \citep{redi82} or perpendicular to neutral planes in the sense of \eqn{neutralvec}  \citep{grif-et-al98}.} \ch{The diffusion process represented by \eqn{redi}} is variously referred to as `rotated', `Redi' or `iso-neutral' diffusion. There is a simple relation between $\tilde{\mathcal{L}}$ and $\mathcal{L}$, namely
\beq
\tilde{\mathcal{L}} = \mathcal{L} + \bm{w} \cdot \nabla, \lab{Lw}
\eeq
where $\bm{w} = - \kappa \bm{n} \cdot \nabla \bm{n}$ is a drift velocity (see Appendix \ref{app:generator} for a derivation). \ch{Therefore, the Stratonovich and It\^o forms of the SDE associated with $\tilde{\mathcal{L}}$ are
\begin{subequations}
\begin{align}
\d \bX &= - \kappa \bm{n} \cdot \nabla \bm{n} \, \d t + \sqrt{2 \kappa} \,  \bm{n} \times \circ \, \d \bm{W} \\
&= -\kappa \nabla \cdot (\bm{n} \otimes \bm{n})  \, \d t + \sqrt{2 \kappa} \, \bm{n} \times  \d \bm{W}.
\end{align}
\end{subequations}
}
Because $\bm{w} \cdot \bm{n} = 0$, $\bm{w}$ can be expressed as a linear combination of the $\bm{v}_k$ and the condition for hypoellipticity of $\mathcal{L}$ and $\tilde{\mathcal{L}}$ is the same, namely the non-integrability condition $\mathcal{H} \not= 0$. 

The processes represented by $\mathcal{L}$ and $\tilde{\mathcal{L}}$ are nonetheless different. In particular, while $\tilde{\mathcal{L}}^* = \tilde{\mathcal{L}}$ leaves the uniform distribution $p = \mathrm{const}$ invariant, \ch{  $\mathcal{L}^*$ does not, since $\tilde{\mathcal{L}}^* \,  1 =0$ but $\mathcal{L}^* \,  1 = \nabla \cdot \bm{w} \not= 0$.} 
In the integrable case, when $\mathcal{H} = 0$ and \eqn{intnormal} holds, $\mathcal{L}^*$ leaves invariant  distributions that are uniform on surfaces $f = \mathrm{const}$ whereas  $\tilde{\mathcal{L}}^*$ does not. We discuss this further in  Appendix \ref{app:generator}. 
In what follows we concentrate on the drift-free model \eqn{smodel1} represented by $\mathcal{L}$. We emphasise that the key distinction between the types of degenerate diffusion -- hypoelliptic or not, depending on whether $\mathcal{H} \not= 0$ or $\mathcal{H} = 0$ -- applies equally to both $\mathcal{L}$ and $\tilde{\mathcal{L}}$. \ch{We focus on $\mathcal{L}$ in what follows because of the simplicity of the associated SDE \eqn{smodel1} even though it does not satisfy the desirable `well-mixed condition' of  \citet{Thomson1987}, namely $\mathcal{L}^* \, 1 \not= 0$.}

Mixing parameterisations often include a weak dianeutral diffusion, leading to a non-degenerate, highly anisotropic diffusion. The qualitative behaviour of degenerate  hypoelliptic and non-degenerate anisotropic diffusions can be sharply different. For short times, the transition probability is roughly
\beq
p(\bx,t|\by) \asymp \e^{-d^2(\bx,\by)/2t},
\eeq
\ch{i.e.\ $\log p(\bx,t|\by) \sim -d^2(\bx,\by)/2t$ as $t \to 0$,} 
for both degenerate and non-degenerate diffusions \citep{vara67,lean87a,lean87b}. In the case of non-degenerate diffusion, the square distance $d^2(\bx,\by)$ is just the weighted Euclidean distance $(\bx-\by) \cdot \mathsf{K} \cdot (\bx - \by)$, with $\mathsf{K}$ the non-singular diffusivity tensor. In the degenerate, hypoelliptic case, $d^2(\bx,\by)$ is the square CC distance. Because of the anisotropic scaling of the CC distance, which is proportional to the Euclidean distance in directions parallel to the neutral planes but to its square root in the perpendicular direction \citep{stri86}, the covariance 
\beq
C^{ij} = \E (X^i - \E X^i)  (X^j - \E X^j)
\eeq
of a fluid parcel's position
has two semi-axes that grow like $t$ and the third, aligned in the local dianeutral direction, that grows like $t^2$. \ch{From the scaling properties of the CC distance, the variance in the dianeutral direction, $D^2_\perp$ say, is expected to scale as $\mathcal{H}^2 D^4_\parallel$, with $D^2_\parallel \propto \kappa t$ the variance in the neutral direction, that is, $D^2_\perp \propto \mathcal{H}^2 \kappa^2 t^2$.}
This behaviour is very different from the non-degenerate case where the three semi-axes grow like $t$, \ch{specifically as $\kappa t$ and $\kappa_\perp t$, where $\kappa_\perp \ll \kappa$ is the dianeutral diffusivity. We  speculate that, if $\kappa_\perp$ is sufficiently small, 
linear and quadratic growths of the dianeutral variance can both be manifested, with a cross-over time $t \sim \kappa_\perp / (\kappa \mathcal{H}^2)$.
}
%although at a much smaller rate in the dianeutral direction than in the neutral directions. 

\ch{While there is a sharp distinction between degenerate and non-degenerate diffusions for short times, long-time and large-distance statistics of strictly
neutral motion is similar to those obtained with an anisotropic non-degenerate diffusion. This is demonstrated numerically in the next section.}
%We demonstrate this in Appendix \ref{app:homo} for a toy model of neutral plane distribution.  

\subsection{Numerical results}

We illustrate both short- and long-time neutral transport in the ocean  by solving \eqn{smodel2} numerically using the distribution of neutral planes estimated from WOCE data. We take somewhat arbitrarily the value $\kappa = 10^4$ m$^2$ s$^{-1}$ for the neutral diffusivity. This is meant to capture in a crude way the effect of all quasi-horizontal turbulence which motivates a value on the large side of typical estimates \citep{kloc-et-al12,aber-et-al13,ying-et-al19}. Changing $\kappa$ only changes linearly the time scale of evolution so the results below can easily be re-interpreted if a different value is preferred. 
 
We solve \eqn{smodel2} in spherical coordinates using Heun's method. This is a consistent discretisation of Stratonovich differential equations \citep[e.g.][]{rume82} that does not require the evaluation of gradients of $\bm{n}$. Explicitly, the position $\bX_k = (\phi_k,\theta_k,z_k)$ of a water parcel at time $t_k = k \Delta t$, $k=0, \, 1, \cdots$, where $\Delta t$ is the time step, is computed using the Euler predictor step
\begin{subequations}
\begin{align}
\tilde \phi_{k+1} &= \phi_{k} + \frac{\sqrt{2\kappa}}{a \cos \theta_k}  h^{1j}(\bX_k) \Delta W_j, \\
\tilde \theta_{k+1} &= \theta_{k} + \frac{\sqrt{2\kappa}}{a}  h^{2j}(\bX_k) \Delta W_j, \\
\tilde z_{k+1} &= z_k +\sqrt{2\kappa}  h^{3j}(\bX_k) \Delta W_j,
\end{align}
followed by 
\begin{align}
\phi_{k+1} &= \phi_{k} + \frac{\sqrt{2\kappa}}{2a} \left(\frac{1}{\cos \theta_k}   h^{1j}(\bX_k) \Delta W_j +  \frac{1}{\cos \tilde \theta_k}  h^{1j}(\tilde{\bX}_{k+1}) \Delta W_j \right) \\
\theta_{k+1} &= \theta_{k} + \frac{\sqrt{2\kappa}}{2a}  (h^{2j}(\bX_k) \Delta W_j+h^{2j}(\tilde{\bX}_{k+1}) \Delta W_j), \\
z_{k+1} &= z_k + \frac{\sqrt{2\kappa}}{2} (h^{3j}(\bX_k) \Delta W_j+h^{3j}(\tilde{\bX}_{k+1}) \Delta W_j),
\end{align}
\end{subequations}
where the $\Delta W_j \sim \mathcal{N}(0,2 \kappa \Delta t)$ are independent Gaussian increments. As in \S\ref{sec:CC} we use a tri-linear interpolation to approximate $\bm{n}(\bX_k)$ and  $\bm{n}(\tilde{\bX}_{k+1})$ based on the gridded values of $\bm{n}$.

\begin{figure}
\begin{center}
\includegraphics[height=5cm]{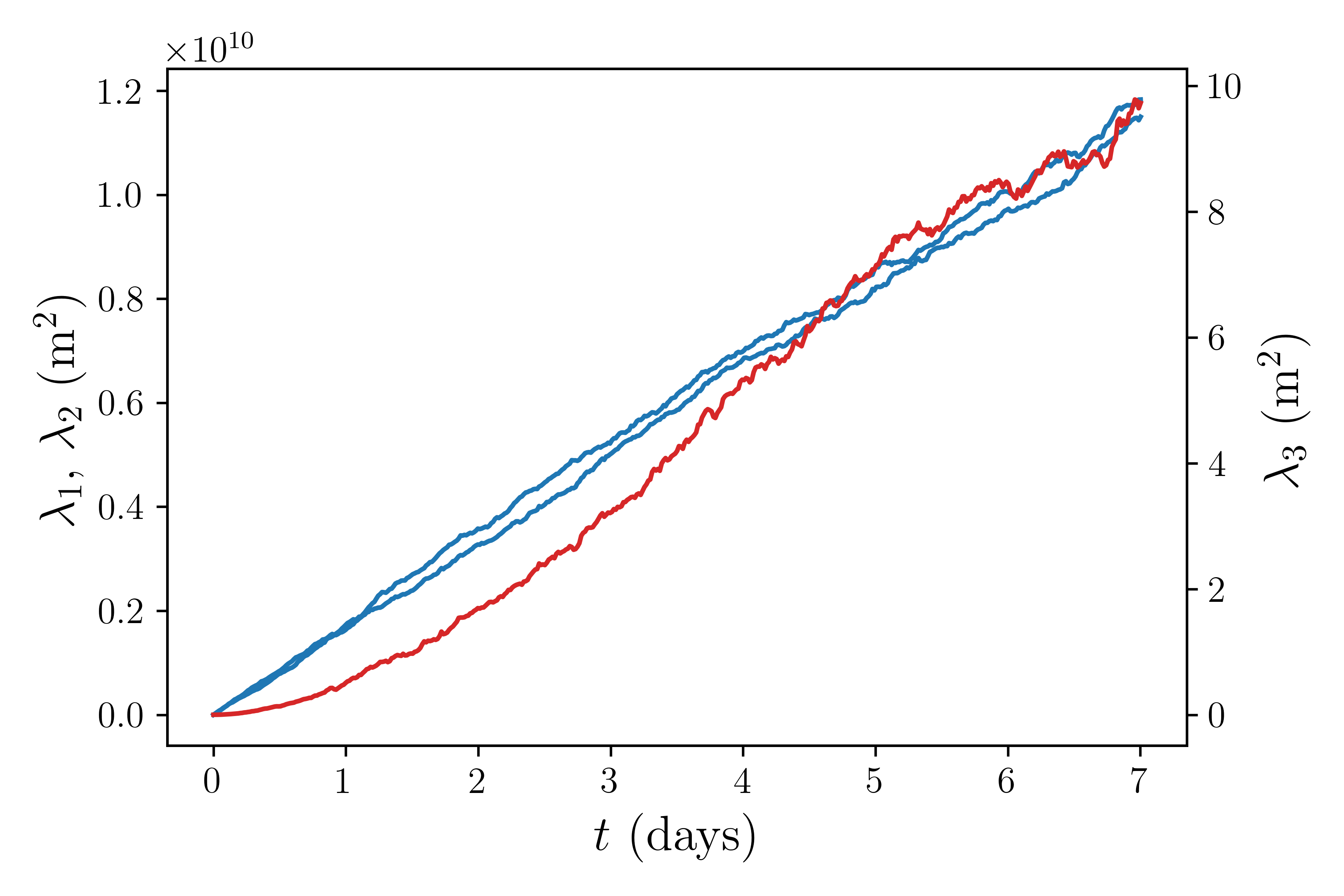} \includegraphics[height=5cm]{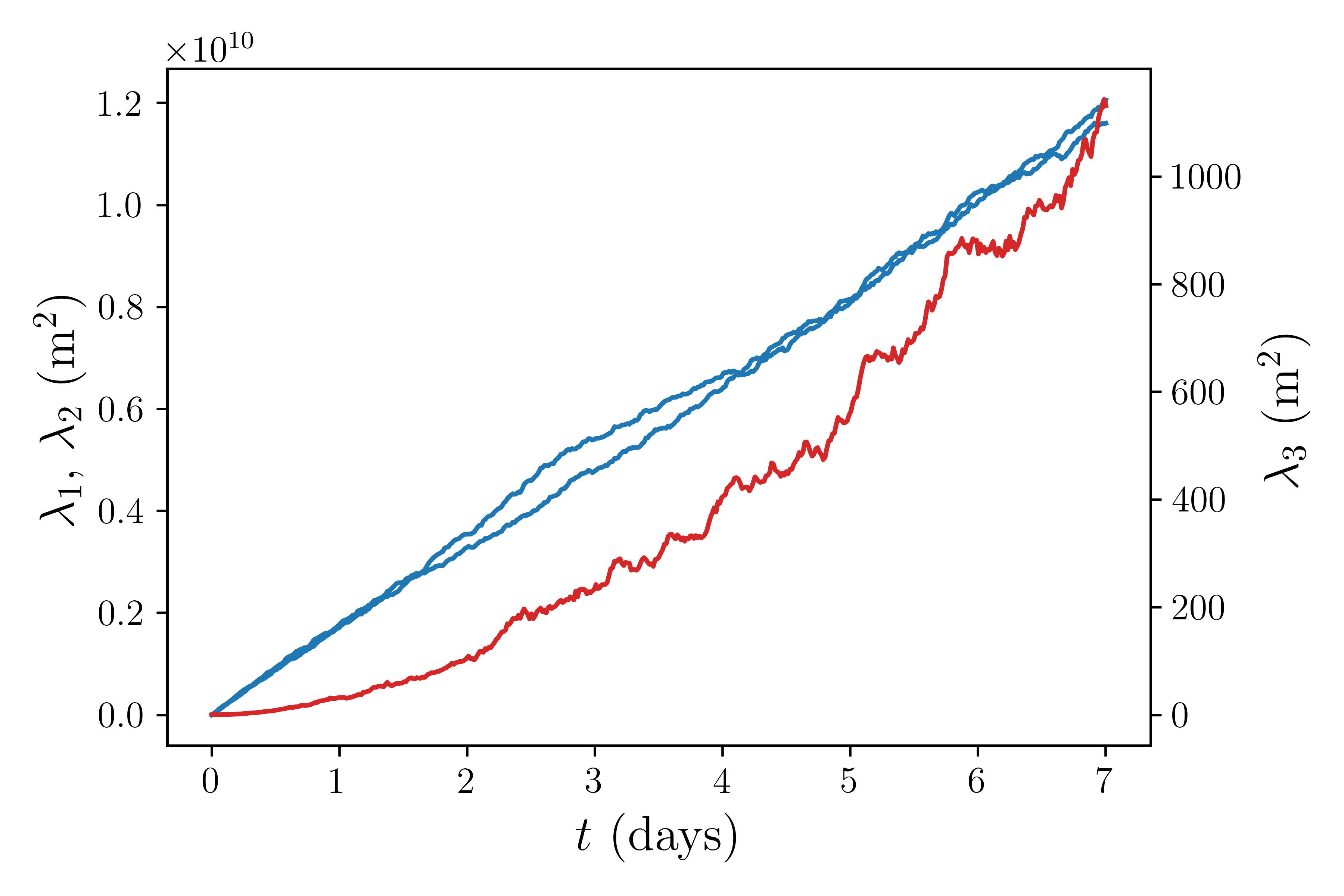}
\caption{Eigenvalues $\lambda_i$ of the covariance matrix $C^{ij}$ of the position of a parcel evolving according to \eqn{smodel1} with $\kappa = 10^4$ m$^2$ s$^{-1}$. The initial position is: longitude $\phi = -25^\circ$, latitude $\theta = 0^\circ$ and depth $z = -2000$ m (left panel), and longitude $\phi = -25^\circ$, latitude $\theta = -50^\circ$ and depth $z = - 2000$ m (right panel). The blue curves, with axis on the left, show the two largest eigenvalues; the red curve, with axis on the right, show the smallest eigenvalues.}
\label{fig:evalues}
\end{center}
\end{figure}
 
Figure \ref{fig:evalues} illustrates the short-time dispersion induced by neutral transport. It shows the evolution of the eigenvalues $\lambda_i, \, i=1,\, 2, \, 3$, of the covariance matrix $C^{ij}$ for 7 days for a parcel starting in the Atlantic Ocean  at a depth of 2000 m at the equator (left panel) and in the Southern Ocean (right panel). The time step is chosen so that the standard deviation of the neutral displacements is $\sqrt{2 \kappa \Delta t} = 5$ km. \ch{This is sufficiently small to make the results insensitive to $\Delta t$.}

As anticipated, the smallest eigenvalue, associated with dianeutral dispersion is much smaller than the other two eigenvalues (which are approximately $2 \kappa t$). Typical displacements over the 7 days are of the order of $100$ km in the approximately neutral directions but only $3$ m or so  in the dianeutral direction near the equator, and $30$ m or so in the Southern Ocean. The difference between dianeutral displacements depending on initial latitude is consistent with the large differences in helicity shown in Figure \ref{fig:Helicity}. The smallest eigenvalue is initially quadratic in $t$ rather than linear, making the unusual, hypoelliptic nature of the diffusion plain. \ch{After this initial phase, all eigenvalues are growing linearly, as can be expected from the central limit theorem and the decorrelation of the neutral directions experienced by parcels.}

\begin{figure}
\begin{center}
\hspace{-4cm}
\includegraphics[height=8cm]{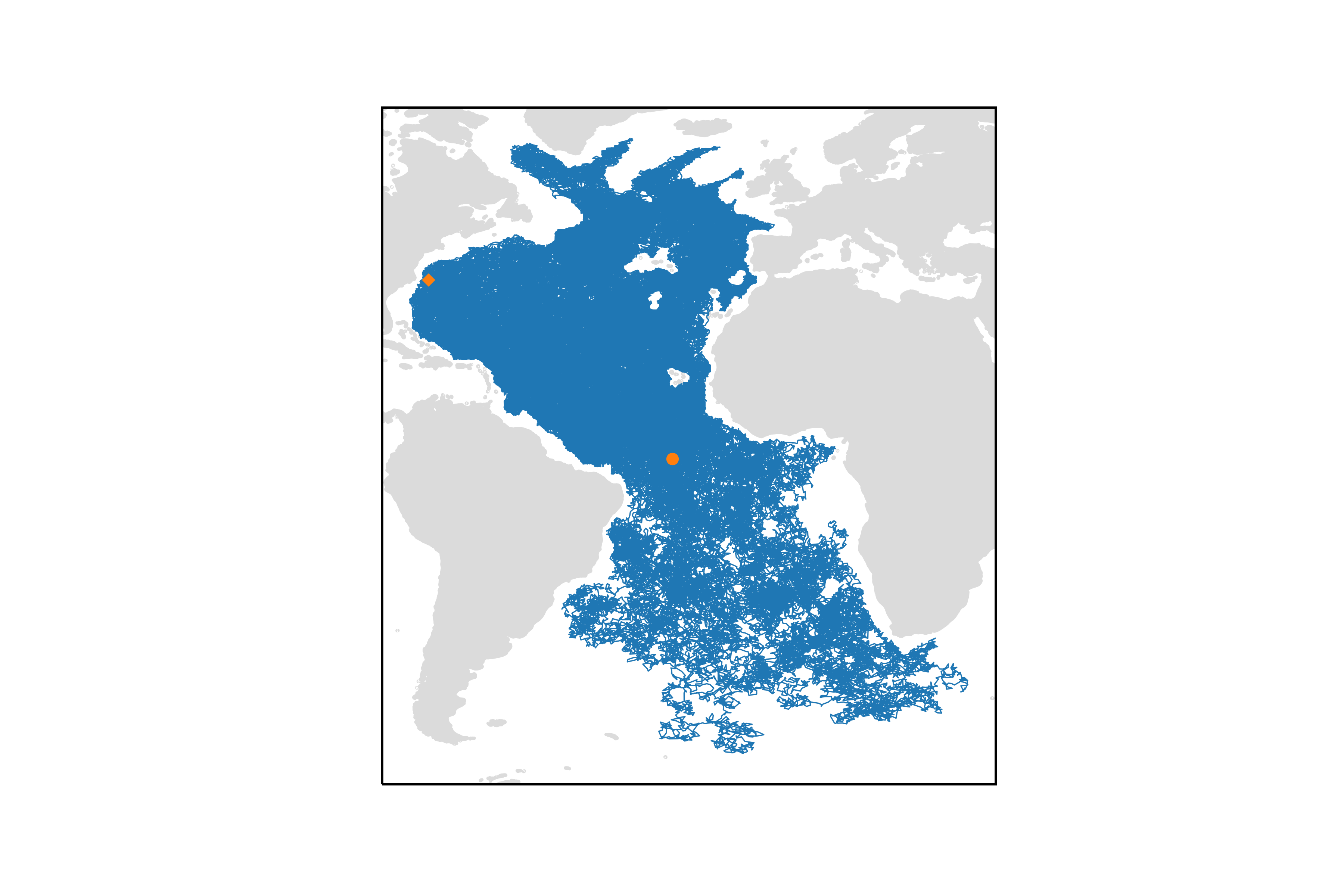} \hspace{-2.5cm} \raisebox{1cm}{\includegraphics[height=6cm]{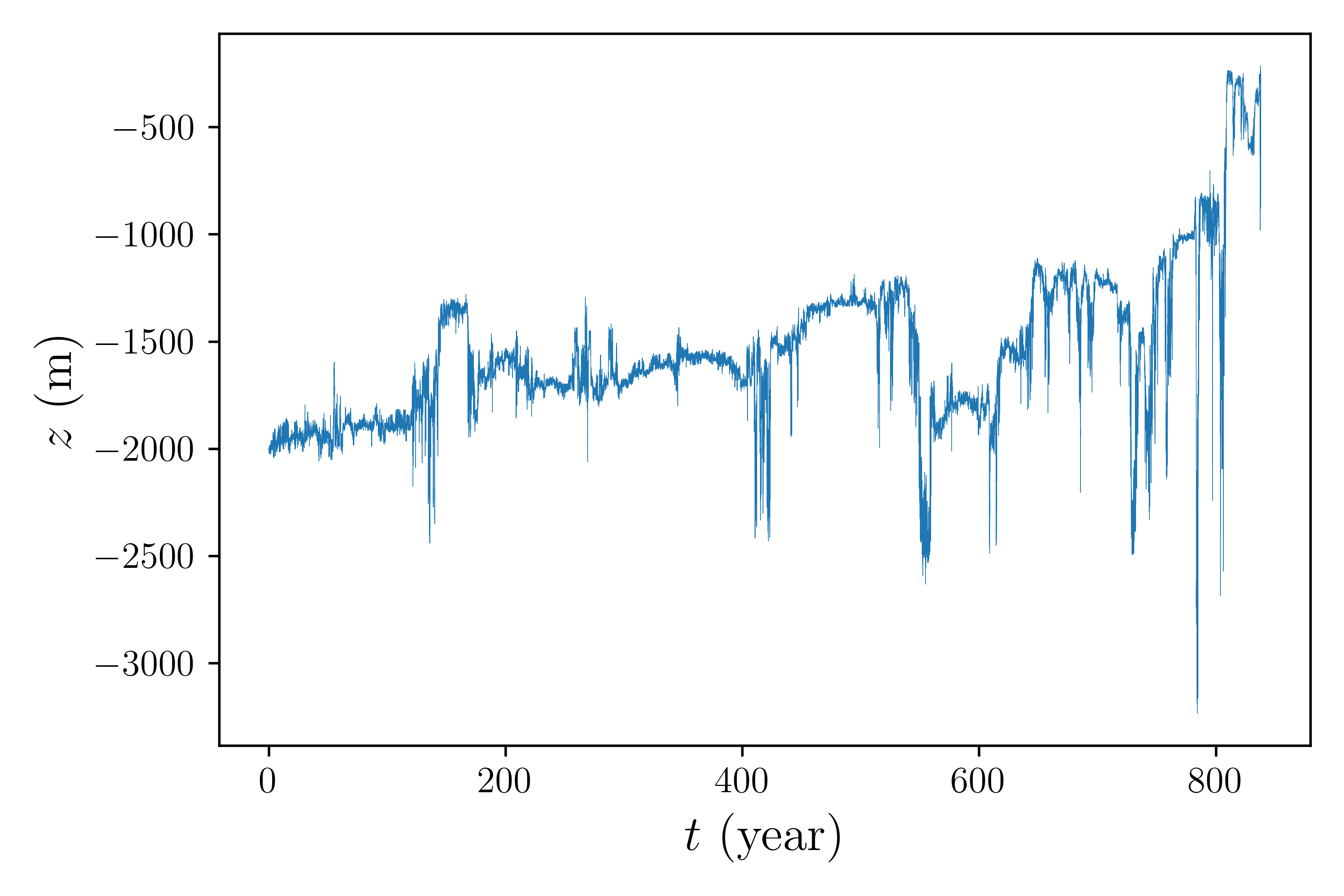}}
\caption{Horizontal trajectory (left) and depth as a function of time (right) for a typical solution of \eqn{smodel2} with $\kappa = 10^4$ m$^2$ s$^{-1}$. The initial position, with longitude $\phi = -25^\circ$, latitude $\theta = 0^\circ$ and depth $z = -2000$ m, is indicated by the orange circle; the final position corresponds to the first passage at $z=-200$ m and is indicated by the orange diamond.}
\label{fig:path}
\end{center}
\end{figure}

Figure \ref{fig:path} illustrates long-time dispersion. It shows the horizontal projection of a single solution of \eqn{smodel2} (left panel) and the corresponding depth $z$ as a function of time (right panel). 
The solution is started at  $\phi = -25^\circ$,  $\theta = -50^\circ$ and  $z = - 2000$ m and stops the first time $z$ reaches $-200$ m, taken as a representative depth of the mixed layer. The time step is such that  $\sqrt{2 \kappa \Delta t} = 50$ km. \ch{We have checked that the resulting statistics are largely insenstive to changes of time step.}

We deal with the presence of boundaries  \ch{by implementing the straightforward `rejection' algorithm. This leaves} the particle at position $\tilde{\bm{X}}_k$ whenever $\tilde{\bm{X}}_{k+1}$ is beyond the boundary of the ocean. \ch{We expect this to match the no-flux behaviour of more sophisticated reflection algorithms in the limit $\Delta t \to 0$  away from boundary layer.}

For the particular realisation of the Wiener process in  \eqn{smodel2} shown in figure \ref{fig:path}, the parcel reaches the mixed layer after about 850 years. In this time, it has sampled much of the Atlantic Ocean, travelling about $10^7$ km in the horizontal. This is typical of the results we obtain for an ensemble of realisations of the Brownian motion.

\begin{figure}
\begin{center}
\includegraphics[height=5.1cm]{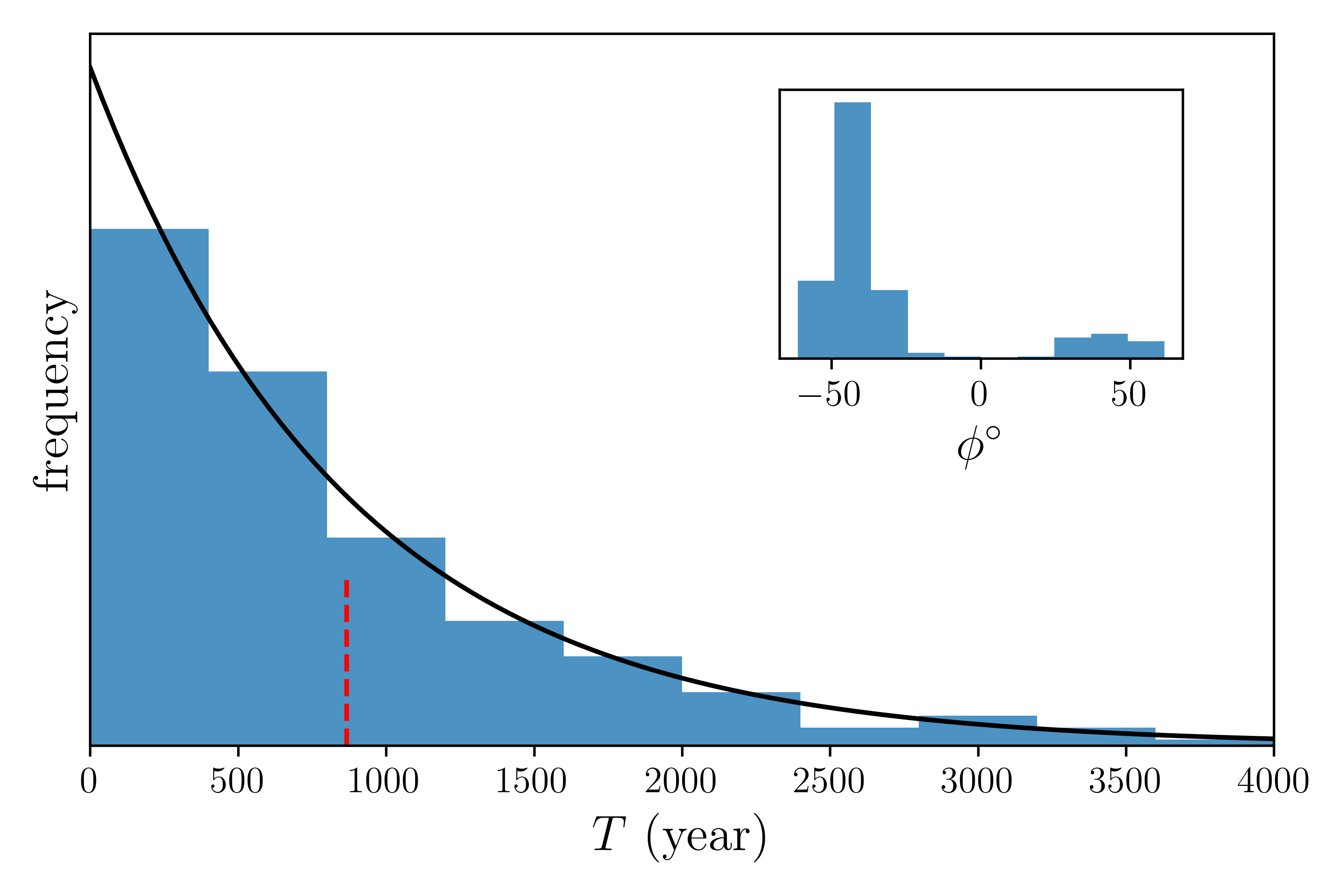}
\includegraphics[height=5.1cm]{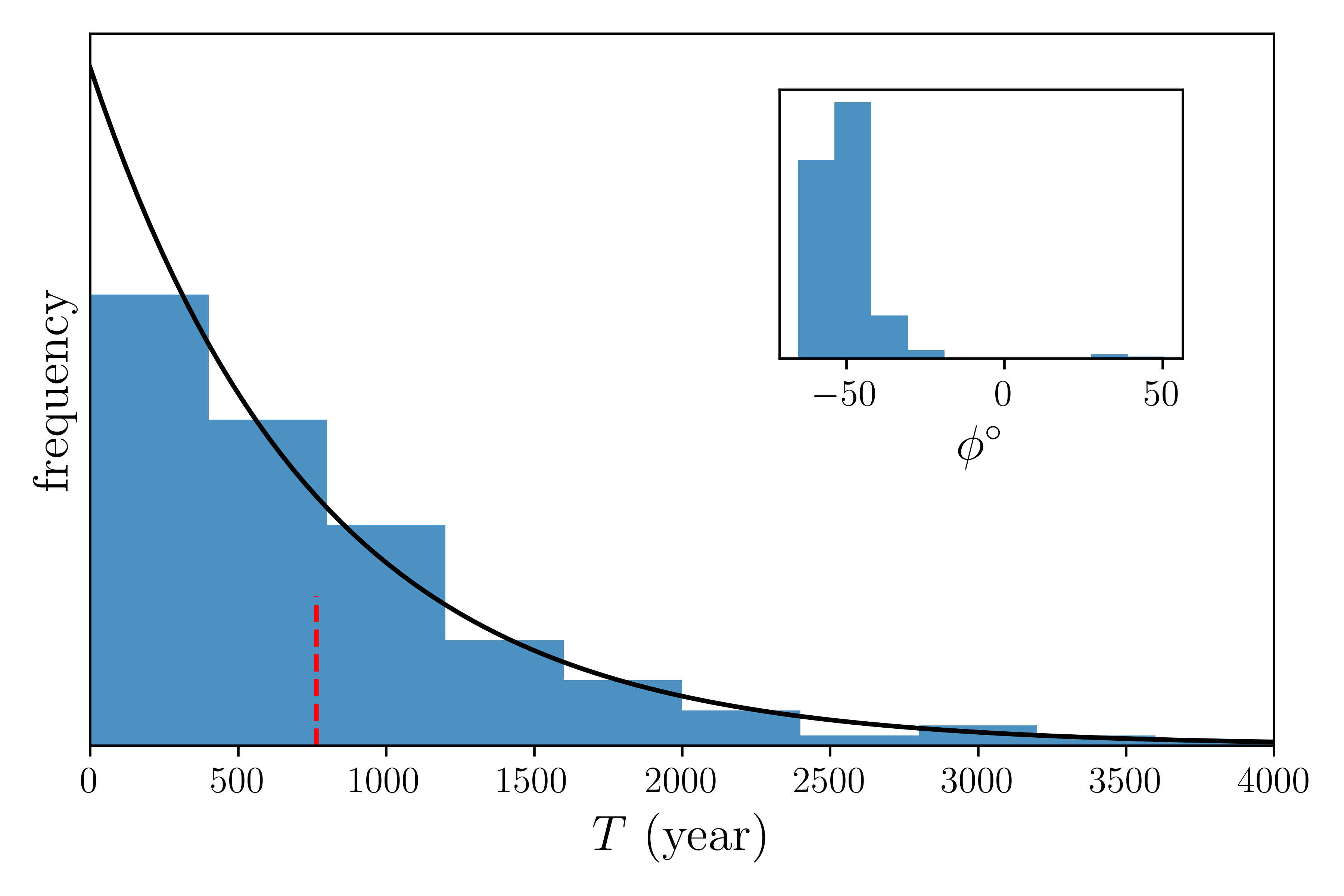}
\caption{Distributions of the time $T$ of first passage in the mixed layer (main panels) and of the latitude of first passage (insets) for parcels moving according to \eqn{smodel2} with $\kappa = 10^4$ m$^2$ s$^{-1}$. The parcels start at latitude $\theta = 0^\circ$, depth $z = -2000$ m and longitude $\phi = -25^\circ$ (left panel, representative of the Atlantic Ocean) or $\phi = -150^\circ$ (right panel, representative of the Pacific Ocean). The distribution of first passage time is well approximated by an exponential distribution (black solid line) with means indicated by the red dashed vertical lines at $\bar T = 866$ years in the Atlantic and $\bar T = 763$ years in the Pacific.}
\label{fig:stats}
\end{center}
\end{figure}

To gain insight into the variability of neutral trajectories, we computed two ensembles of 250 trajectories. For the first ensemble, the trajectories start at the same point in the Atlantic Ocean as for figure \ref{fig:path} ($\phi = -25^\circ$, $\theta = 0^\circ$, $z = -2000$ m). For the second ensemble, the trajectories start at $\phi = -150^\circ$, $\theta = 0^\circ$, $z = -2000$ m, in the Pacific Ocean. The distributions of the time $T$ of first passage in the mixed layer are shown in figure \ref{fig:stats}, together with the distributions of the latitude of the first passage (insets). We fit exponential distributions for $T$ to estimate the mean first passage time $\bar T$. We find that $\bar T \approx 866$ years in the Atlantic and $\bar T \approx 763$ years in the Pacific.

To put these numbers in perspective, we can compare them with the age of tracers in similar areas of the Atlantic and Pacific. Estimates for these are of the order of $300$ years in the Atlantic and $1200$ years in the Pacific according to both radiocarbon data \citep[e.g.][figure 7]{gebb-huyb12} and numerical simulations \citep[e.g.][figure 6]{peac-malt06}. Tracer age quantifies downwelling rather than upwelling but for simplicity we assume that the two processes have similar time scales. The mean first passage times we find under the assumption that the transport is strictly neutral are of a similar order of magnitude to these time scales. 
This suggests that the constraint of neutrality can play a role in setting the time scale of transport from the deep ocean to the surface. Because our model is crude and the first passage times depend (linearly) on an asssumed value for the diffusivity $\kappa$, it would be unreasonable to draw stronger conclusions. 

The distributions of the latitude of first passage in the mixed layer, shown as insets in figure \ref{fig:stats}, are dominated by a sharp peak in the Southern Ocean, at latitudes around $-50^\circ$. This is consistent with the geography of approximately neutral surfaces and with large local values of the helicity. In the Atlantic, there is a significant peak at high latitudes in the Northern hemisphere.

% For comparison, the mean first passage time of a vertical Brownian diffusion for a particle starting at $z$ in an ocean of depth $H$ is
% \beq
% T = - \frac{z(z+2H)}{2 \kappa_{\mathrm{v}}}.
% \eeq
% Here $\kappa_{\mathrm{v}}$ is the vertical diffusivity. If we take its value to be  $\kappa_{\mathrm{v}} = 10^{-4}$ m$^2$ s$^{1}$ chosen as representative of the entire ocean according to Munk's abyssal recipe, we find, for $H=4$ km and $z=-2$ km, $T=10^{10} \, \mathrm{s} \approx 2000$ years.

\section{Discussion}

This paper considers the geometry of neutral paths in the ocean, connecting oceanographic questions with the well developed mathematics of contact and sub-Riemannian geometry. We do not discuss the extent to which parcel trajectories can be treated as neutral. Rather, we take this as a premise and explore the consequences for transport of the neutrality constraint. Our viewpoint is that the strong two-dimensionality of ocean motion can usefully be  described as resulting from motion along a non-integrable distribution of planes, irrespective of the specifics of this distribution. The energetics arguments put forward in favour of specific choices of neutral planes are heuristic \citep{mcdo87,nyca11,tail16b,tail23}. It would of course be preferable to have a first-principle derivation based on approximating the dynamical equations. Such a derivation remains elusive, however.  

Sub-Riemannian geometry gives insights into neutral transport at both local and global levels. At a local level, dianeutral transport can be thought of as resulting from the non-commutation of neutral vector fields. The strong anisotropy of neutral transport can be identified with the strong anisotropy of small sub-Riemannian balls which have sizes $O(t)$ in the neutral directions and $O(t^2)$ in  the dianeutral direction. The anisotropy is exacerbated in the ocean context by the smallness of the helicity. 
At a global level, the CC distance and the corresponding geodesics provide a new perspective on the geometry of the ocean. The ratio of horizontal size of the ocean to the much larger CC depth 
offers an unconventional quantification of the anisotropy of ocean transport. It would be valuable to develop the optimisation tools required for the computation of the CC depth. 

The stochastic model developed in \S\ref{sec:stochastic} illustrates generic properties of neutral transport with small helicity. While motion is not confined to surfaces, since there are no exactly neutral surfaces, it is primarily two-dimensional, diffusing only slowly in the dianeutral direction. There is a large time-scale separation between quasi-neutral and dianeutral motion which could be exploited to obtain asymptotic results. Preliminary results obtained for toy models suggest that over long time scales the dianeutral motion is approximately a one-dimensional diffusion with $O(\mathcal{H}^2)$ diffusivity.  We leave the detailed asymptotic analysis of the stochastic model for future work.

\medskip
\noindent 
\textbf{Acknowledgments.} R\'emi Tailleux and Gabriel Wolf kindly provided the data for the dianeutral field $\bm{n}$ computed from WOCE climatology. We thank R\'emi Tailleux, Paola Cessi and the anonymous referees for valuable comments.
\smallskip

\noindent
\textbf{Disclosure statement.} The authors report there are no competing interests to declare.

\appendix

\section{An interpretation of neutral transport} \label{app:interpr}

It is not clear to which extent water-parcel trajectories are neutral. In particular, neutrality is incompatible with adiabaticity, which requires $\bu \cdot \nabla \theta = \bu \cdot \nabla S = 0$, unless the equation of state is linear. This incompatibility arises because the neutrality condition is derived from a purely local argument that does not extend to finite parcel displacements. The finite-displacement extension of the density-matching condition \eqn{constdens}, namely
\beq
 R\left(\theta(\bx),S(\bx),p(\bm{\varphi}_t(\bx))\right) = R\left(\theta(\bm{\varphi}_t(\bx)),S(\bm{\varphi}_t(\bx))),p( \bm{\varphi}_t(\bx))\right),
\eeq
where $\bm{\varphi}_t(\bx)$ denotes the position at time $t$ of the parcel starting at $\bx$ at $t=0$, cannot hold in general.
It is however possible to recover the neutrality constraint \eqn{perp}--\eqn{neutralvec} from a global argument provided that adiabaticity is replaced by relaxation of potential temperature and salinity to the local conditions, as we now show. 

Suppose that the potential temperature $\vartheta(t)$ and salinity $\mathcal{S}(t)$ of a fluid parcel identified by its starting position $\bx$ obey
\beq
\dot \vartheta = \tau^{-1} \left(\theta(\bm{\varphi}_t(\bx)) - \vartheta \right) \quad \textrm{and} \quad
\dot{\mathcal{S}} =\tau^{-1} \left(S(\bm{\varphi}_t(\bx)) - \mathcal{S} \right),
\lab{relax}
\eeq
where $\tau$ is a relaxation time. Eq.\ \eqn{relax} parameterises 
the effect of small-scale mixing as a simple linear relaxation.  For small $\tau$,  \eqn{relax} can be solved perturbatively as a power series in $\tau$ to find
\beq
\vartheta = \theta - \tau \bu \cdot \nabla \theta +O(\tau^2) \quad \textrm{and} \quad \mathcal{S} = S - \tau \bu \cdot \nabla S +O(\tau^2),
\lab{relaxapp}
\eeq
where the right-hand sides are evaluated at $\bm{\varphi}_t(\bx)$. The matching of the density of the fluid parcel with potential temperature $\vartheta$ and salinity $\mathcal{S}$ with the density of its environment at 
$\bm{\varphi}_t(\bx)$ reads
\beq
R\left(\vartheta,\mathcal{S},p(\bm{\varphi}_t(\bx))\right) = R\left(\theta(\bm{\varphi}_t(\bx)),S(\bm{\varphi}_t(\bx)),p( \bm{\varphi}_t(\bx))\right).
\lab{match}
\eeq
Introducing the approximation \eqn{relaxapp} into \eqn{match} and retaining the leading-order $O(\tau)$ terms recovers the neutrality condition \eqn{perp} with dianeutral vector \eqn{neutralvec}. 

The simple model just described can be thought of as an extension of the original local argument of \citet{mcdo87}. \ch{The model is heuristic: the linear relaxation \eqn{relax} is a crude representation of mixing, one which does not, for instance, guarantee the entropy increase that the second law of thermodynamics requires.
Its aim is to show} that small-scale mixing is integral to the hypothesis that finite-length particle trajectories can be approximately neutral. Different trajectories are obtained if instead of the relaxation to the local environment \eqn{relaxapp}, parcels are assumed to conserve their initial potential temperature and salinity, as proposed by  
\citet{mcdo87b} for submesoscale coherent vortices.

\section{Helicity and Christoffel symbols} \label{app:chrisoffel}

Let $\bv_1, \ \bv_2 \in \Delta$ be generated by the two 1-forms $\bm{p}$ and $\bm{q}$ in the sense that
\beq
\bv_1 = h(\bm{p}), \quad \bv_2 = h(\bm{q}).
\eeq 
Then,
\begin{align}
\boeta([\bv_1,\bv_2]) &= \left( h^{li} p_i \partial_l h^{kj} q_j - h^{lj} q_j \partial_l h^{ki} p_i \right) \eta_k \nonumber \\
&= - 2 \Gamma^{ijk} p_i q_j \eta_k
\end{align}
since $h^{kl} \eta_k=0$. We can relate the left-hand side with \eqn{helicityin} by noting that $\mu(\bv_1,\bv_2,\bm{n}) = \mathrm{vol}(\bm{p},\bm{q},\bm{n}) = \eps^{ijk} p_i q_j \eta_k$, on treating $\bm{p}$ and $\bm{q}$ as vector, with $\bm{v}_1$ and $\bm{v}_2$ as their orthogonal projection on $\Delta$, and
using that $n^k = \eta_k$. This gives
\beq
\mathcal{H} \, \eps^{ijk} p_i q_j \eta_k =  2 \Gamma^{ijk} p_i q_j \eta_k.
\eeq
Since $\bm{p}$ and $\bm{q}$ are arbitrary we conclude that
\beq
 \Gamma^{ijk} \,  \eta_k = \tfrac{1}{2} \mathcal{H} \, \eps^{ijk} \, \eta_k.
 \lab{GammaH}
\eeq
The integrability condition $\mathcal{H}=0$ becomes $\Gamma^{ijk} \,  \eta_k = 0$ in terms of the Christoffel symbols.
This is consistent with the general condition given by \citet{stri86}. This states that non-integrability is equivalent to the injectivity of the map 
\beq
\Gamma^{ijk} \eta_k: \ T_*M /N \to \Delta \quad \textrm{with} \quad \bm{q} \mapsto \Gamma^{ijk} q_j \eta_k,
\eeq
where
$N = \{\alpha \boeta, \, \alpha \in \mathbb{R}\}$.  
%(The sets make sense: for $\bm{p} = \boeta$, $\Gamma^{ijk} q_j \eta_k  \eta_i = 0$, so the image of the map is $\Delta$, and for $\bm{q} = \boeta$,  $\Gamma^{ijk} \eta_j  \eta_k  = 0$ so adding a multiple of $\boeta$ to $\bm{q}$ has no effect.) 
Injectivity means that $\Gamma^{ijk} \,  \eta_k$ has only multiples of $\boeta$ in its null space, which is true for $\mathcal{H} \not= 0$ since the matrix $\eps^{ijk} \eta_k$ has only this null space.

\section{Local form of geodesics} \label{app:local}

We obtain a small-$t$ expression for geodesics using that $\bx(t) = \e^{-t \{H,\cdot\}} \bx(0)$, with $\{\cdot,\cdot\}$  the Poisson bracket. This gives
\beq
x^i(t) = x^i(0) + t \, h^{ij}(0) \, p_j(0) + \tfrac{1}{2} t^2 \, \Gamma^{ijk}(0) \, p_j(0) p_k(0) + O(t^3),
\lab{smallt}
\eeq
where $h^{ij}(0)$ is short for $h^{ij}(\bx(0),\bm{p}(0))$ and similarly for $\Gamma^{ijk}(0)$. This shows that geodesics travel an $O(t)$ (Riemannian) distance in neutral directions (along $\Delta$) and an $O(t^2)$ distance in the dianeutral direction (perpendicular to $\Delta$). We can be more specific: let us take coordinates such that $x^i(0)=0$ and $\eta(0)=d x^3$, that is, $n^1(0)=n^2(0)=0$ and $n^3(0)=1$, so that the dianeutral direction is that of $x^3$. It follows that
\beq
h(0) = \begin{pmatrix} 1 & 0 & 0 \\ 0 & 1 & 0 \\ 0 & 0 & 0 \end{pmatrix}
\eeq
and from \eqn{GammaH} that
\beq
\Gamma^{ij3}(0) = \tfrac{1}{2} \mathcal{H}(0) \, \eps^{ij3}.
\eeq
Eq.\ \eqn{smallt} therefore reduces to
\begin{align}
x^1(t) &= t p_1(0) + \tfrac{1}{2} t^2 \left( \sum_{j,k=1}^2 \Gamma^{1jk}(0) p_j(0) p_k(0) + \mathcal{H}(0) p_2(0) p_3(0)\right), \nonumber \\
x^2(t) &= t p_2(0) + \tfrac{1}{2} t^2 \left( \sum_{j,k=1}^2 \Gamma^{2jk}(0) p_j(0) p_k(0) - \mathcal{H}(0) p_1(0) p_3(0) \right),  \\
x^3(t) &= \tfrac{1}{2} t^2 \sum_{j,k=1}^2 \Gamma^{3jk}(0) p_j(0) p_k(0). \nonumber
\end{align}
This show how, for small helicity $\mathcal{H}(0)$, a geodesic spray, corresponding to $0 \le t \le 1$ and $\bm{p}$ on the unit sphere $|\bm{p}|=1$, is confined near the surface defined by $p_3=0$, separating only by an $O(\mathcal{H})$ distance in the $(x^1,x^2)$ plane for $t=1$.

\section{Generators} \label{app:generator}

We obtain an expression for the generator associated with \eqn{smodel1}. Writing the components of the vector fields $\bv_k$, $k=1,\, 2, \,3$, as
\beq
v_k^i = \sqrt{2\kappa} \, \eps^{ijk} n^j, 
\eeq
we have
\begin{align}
\kappa^{-1} \mathcal{L} &= (2 \kappa)^{-1} v_k^i \partial_i (  v_k^j \partial_j) = \eps^{ilk} \eps^{jmk} n^l \partial_i ( n^m  \partial_j) = \delta_{ij} n^l \partial_i ( n^l \partial_j) 
- n^j \partial_i( n^i \partial_j) \nonumber  \\
&= (\delta_{ij} - n^i n^j) \partial_{ij} - \partial_i n^i \, n^j \partial_j = h^{ij} \partial_{ij} - (\nabla \cdot \bm{n}) n^j \partial_j, \lab{gen1}
\end{align}
where we have used $\eps^{ilk} \eps^{jmk} = \delta^{ij} \delta^{lm} - \delta^{im} \delta^{jl}$ and $n^l \partial_i n^l = \frac{1}{2} \partial_i | \bm{n}|^2 = 0$.

We verify that  \eqn{smodel2} has the same generator and hence the same distribution for $\bX(t)$.   In this case,
\beq
v_k^i = \sqrt{2\kappa} \, h^{ik} = \sqrt{2\kappa} \left( \delta^{ik} - n^i n^k \right)
\eeq
and
\begin{align}
\kappa^{-1} \mathcal{L} &= (2 \kappa)^{-1}  v_k^i \partial_i (  v_k^j \partial_j) = (\delta^{ik} - n^i n^k) \partial_i ((\delta^{jk} - n^j n^k) \partial_j)
 \nonumber  \\
 &= (\delta^{ik} - n^i n^k)  (\delta^{jk} - n^j n^k)  \partial_{ij} -  (\delta^{ik} - n^i n^k) \partial_i (n^j n^k) \partial_j \nonumber \\ 
&= (\delta^{ij} - n^i n^j) \partial_{ij} - (\partial_i ( n^i n^j) - n^i \partial_i n^j - n^i n^j n^k \partial_i n^k ) \partial_j  \nonumber \\
&= h^{ij} \partial_{ij} - (\nabla \cdot \bm{n}) n^j \partial_j, \lab{d4}
\end{align}
matching \eqn{gen1}.

We can read off from $\mathcal{L}$ that the Stratonovich to It\^o 
correction is
\beq
\tfrac{1}{2}  v_k^i \partial_i v_k^j \, \d t = - \kappa (\nabla \cdot \bm{n}) \bm{n} \, \d t,
\eeq
leading to \eqn{smodel1Ito}. 

We next verify the form \eqn{Lw} of the generator $\tilde{\mathcal{L}}$ defined in \eqn{redi}. 
We have
\begin{align}
\kappa^{-1} \tilde{\mathcal{L}} &=  \partial_i \left((\delta^{ij} - n^i n^j) \partial_j \right)   \\ 
&= (\delta^{ij} - n^i n^j) \partial_{ij} - \partial_i n^i n^j \partial_j - n^i \partial_i n^j \partial_j \nonumber \\
&= h^{ij} \partial_{ij} - (\nabla \cdot \bm{n}) n^j \partial_j - n^i \partial_i n^j \partial_j = \kappa^{-1} (\mathcal{L} + \bm{w} \cdot \nabla)
\end{align}
with $\bm{w}$ the vector field with components $w^j = - \kappa n^i \partial_i n^j$. Note that $\bm{w} \cdot \bm{n} = 0$ since $n^j \partial_i n^j = \frac{1}{2} \partial_i |\bm{n}|^2 = 0$. 

Finally, we consider the invariant distributions for the processes with generators $\mathcal{L}$ and $\tilde{\mathcal{L}}$, that is, the solutions $p$ of  $\mathcal{L}^* p = 0$ and $\tilde{\mathcal{L}}^* p = 0$. For $\tilde{\mathcal{L}}$, it is clear from $\tilde{\mathcal{L}} = \tilde{\mathcal{L}}^*$ that $p=1$ (or more generally any constant) is a solution. 
This is not the case for $\mathcal{L}$. Taking the adjoint of \eqn{Lw} and manipulating leads to 
\beq
\kappa^{-1} \mathcal{L}^* = \mathcal{L} + 2 \bm{w} \cdot \nabla + (\nabla \cdot \bm{w})
\eeq
so that $\mathcal{L}^* \, 1 \not= 0$ unless $\nabla \cdot \bm{w}=0$. There is, however, an intuitive family of invariant disributions for $\mathcal{L}$ in the integrable case, with $\bm{n} = \nabla f / |\nabla f|$ for some $f$, namely
\beq
p(\bx) = |\nabla f(\bx)| \delta(f(\bx) - c),
\lab{pc}
\eeq
for any constant $c$. This corresponds to a uniform distribution on the surface $f = c$ and is consistent with the interpretation of $\mathcal{L}$ as the generator of Brownian motion on this surface. 

We check the invariance of $p$ in \eqn{pc} for completeness.
We rewrite $\mathcal{L}^*$ as
\beq
\mathcal{L}^* p = \kappa \partial_i ( (\delta_{ik} - n^i n^k) \partial_j ( (\delta_{jk} - n^j n^k) p))
\eeq
and compute the flux
\beq
F^i = \kappa (\delta_{ik} - n^i n^k)  \partial_j ( (\delta_{jk} - n^j n^k) p) = \kappa (\delta_{ij} - n^i n^j) \partial_j p +  w^i p.
\eeq
For $p$ in \eqn{pc} we obtain
\beq
F^i = \left( \kappa \ (\delta_{ij} - n^i n^j) \partial_j |\nabla f| +  w^i |\nabla f| \right) \delta(f-c) = 0.
\eeq
The final cancellation follows from observing that $|\nabla f| = n^j \partial_j f$, applying $\partial_i$ and using that $\partial_i f = |\nabla f| n^i $ to obtain $(\delta_{ij} - n^i n^j) \partial_j |\nabla f| = n^j \partial_j n^i = - \kappa^{-1} w^i$.

\bibliographystyle{gGAF}
\bibliography{mybib}
 
\end{document}